\documentclass[a4paper,fleqn]{cas-sc}
\usepackage{lineno}
\usepackage[authoryear]{natbib}
\bibpunct{(}{)}{;}{a}{,}{,}
\usepackage{hyperref}
\usepackage{multirow}
\usepackage{amsmath}
\usepackage{rotating,float,caption}
\usepackage{setspace}   

\def\tsc#1{\csdef{#1}{\textsc{\lowercase{#1}}\xspace}}
\tsc{WGM}
\tsc{QE}
\tsc{EP}
\tsc{PMS}
\tsc{BEC}
\tsc{DE}

\begin{document}
\let\WriteBookmarks\relax
\def\floatpagepagefraction{1}
\def\textpagefraction{.001}
\shorttitle{Meteor observations at Venus}
\shortauthors{A.~A.~Christou et~al.}

\title [mode = title]{Feasibility of meteor surveying from a Venus orbiter}                      



\author[1]{Apostolos A.~Christou}[]
\cormark[1]
\ead{Apostolos.Christou@armagh.ac.uk}


\address[1]{Armagh Observatory and Planetarium, College Hill, Armagh BT61 9DG, Northern Ireland, United Kingdom}

\author[2,3]{ Maria Gritsevich}



\address[2]{Faculty of Science, Gustav H\"{a}llstr\"{o}min katu 2, FI-00014 University of Helsinki, Finland}
\address[3]{Institute of Physics and Technology, Ural Federal University, Mira str. 19, 620002 Ekaterinburg, Russia}





\cortext[cor1]{Corresponding author}

\doublespacing

\begin{abstract}
Meteor and bolide phenomena caused by the atmospheric ablation of incoming meteoroids are predicted to occur at the planet Venus. Their systematic observation would allow to measure and compare the sub-mm to m meteoroid flux at different locations in the solar system. Using a physical model of atmospheric ablation, we demonstrate that Venus meteors would be brighter, shorter-lived, and appear higher in the atmosphere than Earth meteors.

To investigate the feasibility of meteor detection at Venus from an orbiter, we apply the SWARMS survey simulator tool to sets of plausible meteoroid population parameters, atmospheric models and instrument designs suited to the task, such as the Mini-EUSO camera operational on the ISS since 2019.

We find that such instrumentation would detect meteors at Venus with a 1.5$\times$ to 2.5$\times$ higher rate than at Earth. The estimated Venus-Earth detection ratio remains insensitive to variations in the chosen observation orbit and detector characteristics, implying that a meteor survey from Venus orbit is feasible, though contingent on the availability of suitable algorithms and methods for efficient on-board processing and downlinking of the meteor data to Earth.

We further show that a hypothetical camera onboard the upcoming {\it EnVision} mission to Venus similar to the ISS instrument should detect many times more meteors than needed for an initial characterisation of the large meteoroid population at 0.7 au from the Sun.
\end{abstract}
\doublespacing


\begin{highlights}
\item We use physics-based modelling to understand the properties of meteors in the upper atmosphere of Venus and to compare meteor survey efficiency from Earth orbit and from Venus orbit

\item Venus meteors would be brighter and shorter-lived than Earth meteors, due to different atmospheric density scale heights
\item Assuming similar meteoroid populations at the two planets, orbital meteor surveys would detect 1.5$\times$--2.5$\times$ more meteors per hour at Venus than at Earth
\end{highlights}

\maketitle
\doublespacing

\section{Introduction}
Meteoroids that efficiently ablate in the Earth's atmosphere are broadly grouped into {\it streams} and {\it sporadics} \citep{CampbellBrown2007}. Stream meteoroids were ejected from a parent comet or asteroid relatively recently - typically in the last $10^{2}$--$10^{4}$ yr - and retain a dynamical memory of their birth so that they can be readily associated with a specific parent body \citep{Valsecchi.et.al1999,Gritsevich.et.al2022}. As these stream meteoroids approach Earth, they follow nearly parallel paths and manifest in the atmosphere as meteor showers, with the most intense ones dominating the flux of visible meteors for periods of a few days. Sporadic meteoroids, on the other hand, arrive at the Earth throughout the year from different directions and dominate the annually-averaged mass influx to our planet \citep{Brown.et.al2008}. Streams and sporadics are genetically related, as sporadics represent former stream meteoroids that have evolved to their current orbits through planetary gravitational scattering, solar-thermal radiation forces, and inter-particle collisions \citep{Grun.et.al1985,Wiegert.et.al2009}. 

Observations of meteors hold the potential to characterise the flux of meteor-producing meteoroids - typically 0.1-1000 mm in size - across a range of heliocentric distances by systematically monitoring planetary atmospheres from orbit \citep{Jenniskens2006,Christou.et.al2007b,Christou.et.al2019}. Being the planet with the nearest orbit to the Earth's, Venus naturally lends itself to this task. Theoretical and laboratory studies \citep{McAuliffe2006,McAuliffeChristou2006a,Blaske.et.al2023} suggest that visible meteors readily occur at Venus, the brightest of which may even be detectable from Earth \citep{BeechBrown1995}.
While meteors are typically recorded using ground-based, upward-looking optical cameras, such an approach is impractical at Venus due to its thick, opaque atmosphere and harsh surface conditions. Conversely, the concept of a camera looking down on the planet from orbit is well-suited to meet this challenge.

Recent models of the terrestrial sporadic flux are calibrated against a diversity of datasets obtained at Earth: meteor counts obtained by radar, observations of zodiacal dust emission, meteoric metal abundances in the upper atmosphere or samples of unmelted micrometeorites   \citep{Wiegert.et.al2009,Nesvorny.et.al2010,Nesvorny.et.al2011,CarrilloSanchez.et.al2016}.
Extrapolated to Venus, these models predict a similar overall mass influx to the terrestrial case ($\sim$30 t $\mbox{d}^{-1}$) \citep{Frankland.et.al2017,CarrilloSanchez.et.al2020} as might perhaps be expected from consideration of the planetary sizes and the proximity of the orbits. The models also predict significant differences in the spatial and temporal distribution of the flux during the planetary year, owing to the different orbit shapes and orientations, rotation rates and obliquities of the rotation axis \citep{Janches.et.al2020}. 

Obtaining meteor data at Venus will therefore allow to test models of the meteoroid influx in the inner solar system which are currently heavily-anchored to Earth observations. The same observations will also lead to the discovery of new meteoroid streams and the study by proxy of their cometary parent bodies, will test our understanding of ablation physics under non-terrestrial conditions and also help mitigate the risk of impact-induced damage to spacecraft \citep{ChristouVaubaillon2010,Schimmerohn.et.al2018,Kruger.et.al2021}.
The existence of Venusian meteors is also relevant to the interpretation of short-lived optical transients (``flashes'') observed recently by the Japanese Exploration Agency's {\it Akatsuki} spacecraft \citep{Lorenz.et.al2022,Blaske.et.al2023}. Similar transients have been reported during an earlier ground-based search for lightning flashes \citep{Hansell.et.al1995} while a feature recorded by the Pioneer Venus Orbiter Ultraviolet Spectrometer instrument in 1979 was interpreted as the $\sim$900-km long trail of a grazing meteor \citep{HuestisSlanger1993}.

As {\it Akatsuki} continues its atmospheric monitoring mission, Venus features in the deep space exploration plans of India, the European Space Agency, the United States, China \citep{Widemann.et.al2023} as well as private sector initiatives \citep{Seager.et.al2022} presenting potential opportunities for conducting meteor surveys at Venus. Despite holding clear advantages over ground-based surveys \citep{Bouquet.et.al2014}, the record of spaceborne meteor detection has until now been the result of serendipity, a case in point being the set of 29 Leonids recorded by the US Ballistic Missile Defense Organization Midcourse Space eXperiment (MSX) satellite in 1997 \citep{Jenniskens.et.al2000} when the estimated Zenithal Hourly Rate of that shower \cite[ZHR; the theoretical number of shower meteors seen by an observer under ideal conditions, see e.g.][]{Beech2006} was $\sim$100, 1$-$2 orders of magnitude less than that of the 1999-2002 Leonid storms \cite[with ZHR in the 1000s; e.g.][]{Soja.et.al2015} and more typical of annually-recurring showers such as the August Perseids or December Geminids \citep{Jenniskens1994,Jenniskens.et.al2016}.

This paper takes a quantitative look at the feasibility of observing meteors in the atmosphere of the planet Venus from an orbiter spacecraft.
Given the present level of maturity in the field, it is appropriate to produce quantitative predictions to support future instrument development and to optimize the design of orbital meteor surveys. We focus mainly on the sporadic component of the meteoroid flux which is better constrained from the Earth observations and available models. While Venus-crossing meteoroid streams should exist in abundance at Venus \citep{Beech1998,Christou2004a,Christou2004b,Christou2010}, the characteristics of the meteor showers they produce will generally differ from one stream to the other. We note that successful forecasting of year-on-year variations of shower activity at Earth often relies on the availability of prior observations of the same shower \citep{Asher.et.al1999,McNaughtAsher1999,Vaubaillon.et.al2005a,Vaubaillon.et.al2005b,Vaubaillon.et.al2023} which are not available for Venus. Meteor rate predictions for the unobserved Venus showers will therefore require a case-by-case treatment, taking into account the specific properties of the parent body, the meteoroid production function and the orbital history of the stream. In the last Section of this paper, the contribution of shower meteors to the overall number of detections expected from an orbital platform at Venus is extrapolated from the observational record of terrestrial activity in order to estimate the overall detection rate of a hypothetical camera on-board the future ESA {\it EnVision} orbiter. 

In the following Section, we introduce a theoretical framework to understand the principal differences between terrestrial and Venusian meteors, enhancing previous findings obtained by numerically solving the ablation equations. Our analytical treatment of atmospheric ablation relies on specific approximations in the regime where meteoroid ablation has begun before significant deceleration sets in. This limits the size of a meteoroid treatable by our model, the limit in terms of meteor brightness being roughly equivalent to the faintest fireballs (i.e.~approximately $-4$ absolute magnitude). While the general theory remains valid for larger meteoroid sizes, our survey simulations primarily focus on fainter meteors treatable by the approximate theory. This is because detection statistics are dominated by these fainter meteors, where the relevant mathematical expressions are much simpler than the corresponding exact forms. In Sections 3 \& 4, we complement the analytical results with simulations of meteor surveying by suitable instrument models at Earth and Venus orbit using the sophisticated SWARMS code. Section 5 details the further application of SWARMS to understand how our earlier conclusions depend on instrument characteristics and meteoroid environment parameters. Finally, in Section 6 we summarize our main findings and outline potential avenues for follow-on studies.
\begin{table}[width=0.5\columnwidth,cols=3,pos=ht!]
\caption{Notation used for the principal quantities in this paper.}\label{TBL:Notation}
\centering
\begin{tabular}{lp{2cm}p{4cm}}
\toprule
Symbol & Units & Description \\
\midrule
$\rho_{m}\mbox{,}$ $h\mbox{,}$ $V\mbox{,}$ $M\mbox{,}$ $S$ & kg $\mbox{m}^{-3}\mbox{,}$ km,\par km $\mbox{s}^{-1}\mbox{,}$ g, $\mbox{m}^{2}$ & Meteoroid\,bulk\,density, height, speed, mass and cross-sectional\,area\,during atmospheric entry \\
$\rho_{0}\mbox{,}$ $M_{0}\mbox{,}$ $V_{0}$ & kg $\mbox{m}^{-3}\mbox{,}$ g,\par km $\mbox{s}^{-1}$ & Meteoroid\,pre-atmospheric density, mass and speed \\
$\gamma$ & degrees  & Angle of meteoroid flight path to horizon\\
$\rho_{a}$, $h_{0}$ & kg $\mbox{m}^{-3}$, km  & Atmospheric density and scale height\\
$\rho_{0}$ & kg $\mbox{m}^{-3}$  & Atmospheric density at $h$$=$$0$\\
$\bar{\rho}$$=$$\rho_{a}$$\rho^{-1}_{0}$ & --  & Normalised atmospheric density\\
$\mu$  &  --  &  Meteoroid shape coefficient \\
$H^{\ast}$ & J $\mbox{kg}^{-1}$ &  Effective enthalpy of destruction  \\
$c_{d}$  &  --  & Atmospheric drag coefficient \\
$c_{h}$  &  --  & Heat transfer coefficient \\
$\sigma$$=$$c_{h} {\left(c_{d} H^{\ast}\right)}^{-1}$ &  $\mbox{s}^2$ $\mbox{m}^{-2}$  & Meteoroid ablation coefficient \\
$A_{0}$$=$$S_{0}\rho^{2/3}_{m}M^{-2/3}_{0}$  &  --  & Meteoroid shape factor \\
$E_{kin}$ & J & Meteoroid kinetic energy \\
$I$, $m$  &  W  & Meteor luminous intensity and stellar magnitude \\
$h_{max}$, $I_{max}$ &  km, W  & Meteor height and luminous intensity at maximum brightness\\
$h_{end}$  &  km  & Meteor end height \\
$\tau$, $\tau_{G}$  &  --  & Instantaneous and global luminous efficiency factors \\
\bottomrule
\end{tabular}
\end{table}

\section{Model of the ablation of small meteoroids}
Meteors emit light via collisional excitation of line emission as the ablated meteoric vapour interacts with atmospheric species \citep{Bronshten1983,Ceplecha.et.al1998}. In the 200$-$700 nm spectral region, the strongest spectral lines correspond to meteoric species such as Mg, Fe and Na rather than atmospheric species such as O or N \citep{Millman1963,Bronshten1983,Ceplecha.et.al1998,Carbary.et.al2003}. We therefore expect the meteor integral radiation at these wavelengths to be only weakly dependent on atmospheric composition. Strong near-infrared atmospheric lines such as OI at 777 nm are sometimes found in spectra of fast meteors \cite[so-called ``type Y'' spectra;][]{Ceplecha.et.al1998} however such meteors are rare in the observational record. Therefore, in this paper we assume that meteor light intensity is proportional to the rate of ablative mass loss from the meteoroid and that empirically-derived scalings used to obtain one from the other are atmosphere-independent. The contribution of atmospheric emissions to meteor light and its variation with atmospheric composition is an interesting open question beyond the scope of the present study. 

During atmospheric entry, a meteoroid with mass $M$ and speed $V$ generates a meteor of luminous intensity $I$ given by:
\begin{linenomath*}
\begin{equation}
\label{eq:intensity}
I = -\tau \frac{d E_{\rm kin}}{dt}
\end{equation}
\end{linenomath*}
where the luminous efficiency $\tau$ is the fraction in the instantaneous meteoroid loss rate of kinetic energy $E_{\rm kin}$=0.5$M$$V^{2}$ converted into light. The rate of energy dissipation is evaluated by simultaneously solving for ablative mass loss and deceleration \cite[e.g. ][]{Bronshten1983}
\begin{linenomath*}
\begin{eqnarray}
\label{eq:deceleration}
M \frac{d V}{d t} =-\frac{1}{2}c_{d}\rho_{a}V^{2} S  & & \\
\label{eq:massloss}
H^{\ast} \frac{d M}{d t} = - \frac{1}{2}c_{h}\rho_{a}V^{3} S  & & 
\end{eqnarray}
\end{linenomath*}
where $H^{\ast}$ is the effective enthalpy of destruction, $c_{d}$ and $c_{h}$ are respectively drag and heat transfer coefficients, $\rho_{a}$ is the atmospheric density at the corresponding height, and $S$ is the meteoroid cross section presented to the airflow.

For meteoroids small enough to ablate completely before noticeable deceleration sets in, meteor intensity can be determined from mass loss alone (Eq.~\ref{eq:massloss}) and $\tau$ can be replaced by the so-called global luminous efficiency $\tau_{G}$, the fraction of the initial kinetic energy converted to luminous energy. In the special case of an exponential atmosphere with density scale height $h_{0}$, energy dissipation in Eq.~\ref{eq:intensity} can be expressed analytically, using the height equation $dh/dt = - V \sin{\gamma}$ and the successive transformations $t \rightarrow h \rightarrow \bar{\rho}$=$\exp(-h/h_{0})$ to re-write Eq.~\ref{eq:massloss} as
\begin{linenomath*}
\begin{equation}
\label{eq:masslossheight}
\frac{d M}{d h} =\frac{M_{0} K}{h_{0}(1 - \mu)} \bar{\rho} (1 - K \bar{\rho})^{\mu / (1 - \mu)}
\end{equation}
\end{linenomath*}
where
\begin{linenomath*}
\begin{equation}
\label{eq:Kdef}
K= \frac{(1 - \mu)c_{h} \rho_{0}S_{0}h_{0}V^{2}_{0}}{2 H^{\ast}\sin{\gamma} M_{0}}
\end{equation}
\end{linenomath*}
and $M_{0}$, $S_{0}$ \& $V_{0}$ are the meteoroid's pre-atmospheric mass, cross-section and speed respectively, $\rho_{0}$ is defined through the relationship $\rho_{a}$$=$$\rho_{0}\bar{\rho}$ and $\gamma$ is the flight path angle relative to the horizon. Our analytical treatment is derived from that used previously \citep{Gritsevich2009,Gritsevich2010,GritsevichKoschny2011,2012CosRe..50...56G,Bouquet.et.al2014} as the basis for a least-square-minimisation process to retrieve the meteoroid atmospheric deceleration, mass loss and luminosity history from observations of the meteor flight and determine its outcome. Although we do not consider non-exponential atmospheric profiles in this study, the procedure can also incorporate an arbitrary atmospheric model using the recently introduced atmospheric height correction method \citep{LyytinenGritsevich2016}.

During ablation, the meteoroid cross-section $S$ is assumed to depend on the remaining mass $M$ as $S/S_0$=$(M/M _0)^{\mu}$ where $0$$<$$\mu$$<$$2/3$, the upper limit corresponding to a shape of the body being constant (self-similar) during the flight \citep{Levin1956,Bouquet.et.al2014,Sansom.et.al2019,Moreno.et.al2020}.

The right-hand-side in Eq.~\ref{eq:masslossheight} vanishes at $\bar{\rho}$$=$$0$ and at $\bar{\rho}(h_{end})$$=$$K^{-1}$, $h_{end}$ being the meteor ending height. By differentiating (\ref{eq:masslossheight}) wrt $h$ and expressing the result as a function of $\bar{\rho}$, we obtain the height of {\it maximum light}:
\begin{linenomath*}
\begin{equation}
\label{eq:maxheight}
\bar{\rho}(h_{max})=(1 - \mu) K^{-1}\mbox{.}
\end{equation}
\end{linenomath*}
\begin{figure}
\centering
\includegraphics[width=0.5\columnwidth]{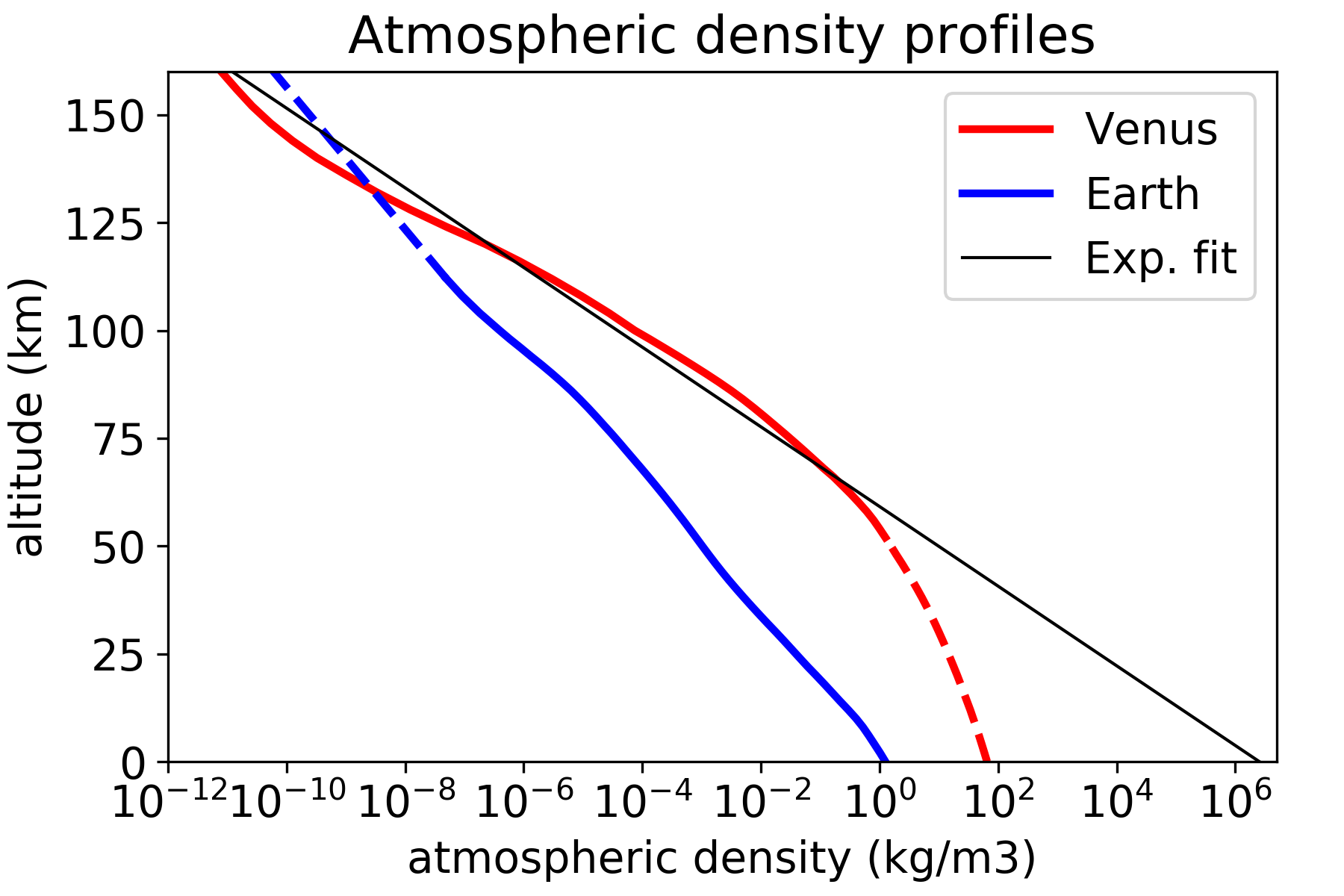}
\caption{Atmospheric density profiles for Earth \citep{Minzner1983} and for Venus \citep{Seiff1983}. The dashed part of the Earth profile represents an extrapolation from the data. The black line represents a fit to the Venus data with the dashed part not considered in the fit.}
\label{FIG:earthvenusprofiles}
\end{figure}
\begin{figure}
\centering
\includegraphics[width=0.5\columnwidth]{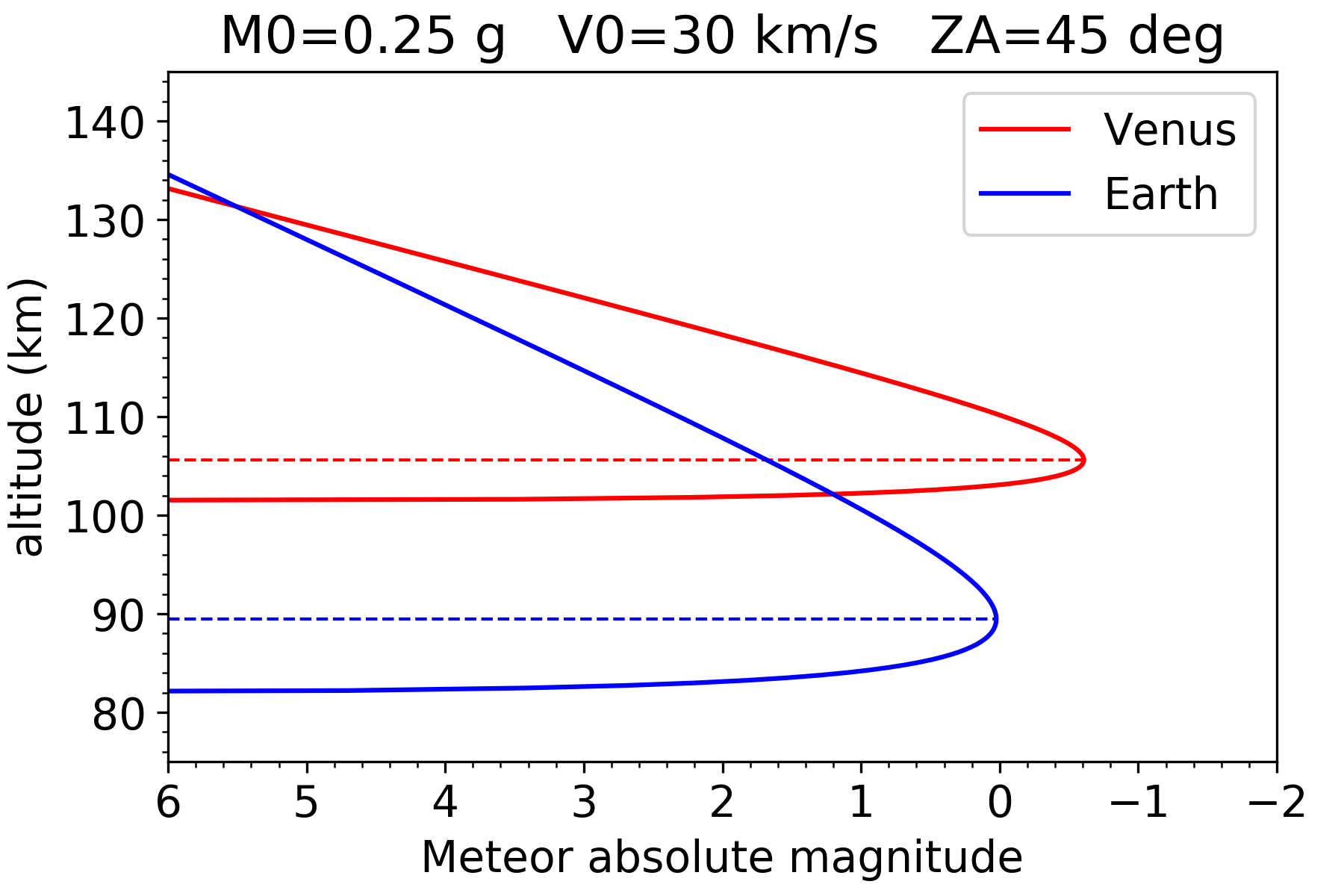}
\caption{Magnitude-height profile for a meteoroid ablating in the atmospheres of Venus and the Earth. The dashed lines indicate the heights of maximum light calculated from theory.}
\label{FIG:earthvenusmeteor}
\end{figure}
Eqs \ref{eq:intensity} \& \ref{eq:massloss} show that, all else being equal, meteor brightness is determined by the atmospheric density $\rho_{a}$. While the Earth’s atmosphere is quasi-exponential from the surface up to $h$$>$$100$ km (Fig.~\ref{FIG:earthvenusprofiles}), at Venus the characteristic scale height decreases from the lower to the middle atmosphere, then remains fairly constant up to $\sim$130 km. Consequently, despite the $\sim$2-orders-of-magnitude Venus/Earth surface density difference, respective density values are similar above $\sim$100 km \citep{Minzner1983,Seiff1983}.

In the Earth’s atmosphere, meteors typically appear at 80--110 km \citep{Jenniskens2006} while at Venus, the same values of atmospheric density are encountered at 105--125 km. To apply the meteor model, we fitted the Venus data from \citet{Seiff1983} to the function $\rho_{a} = \rho_{0} \exp(-h/h_{0})$. The resulting fit (black line) has $h_{0V} = 4.0$ km and $\rho_{0V} = 2.53 \times 10^{6}$ kg $\mbox{m}^{-3}$. For the Earth, we adopt the values \cite[][see also \citet{GritsevichStulov2006} and \citet{Moreno.et.al2015}]{Bouquet.et.al2014}: $h_{0E} = 7.16$ km and $\rho_{0E} =1.29$ kg $\mbox{m}^{-3}$. 

The expression for the dimensionless coefficient $K$ (Eq.~\ref{eq:Kdef}) depends on parameters such as $c_{h}$ and $H^{\ast}$ that are poorly constrained from observations and often hinder the computation process. However, $K$ can also be expressed as $K$=$2\alpha\beta$, where the ballistic coefficient $\alpha$ and mass loss parameter $\beta$ admit to straightforward physical interpretations \citep{Stulov1997,Moreno.et.al2020}. These parameters can be directly derived for each meteor from the observations, eliminating the need for prior knowledge of the poorly constrained parameters \citep{Gritsevich2009,LyytinenGritsevich2016,Sansom.et.al2019,boaca2022characterization}. Given the absence of meteor observations in the Venusian atmosphere for inversion studies and our aim to forward-model the meteor lightcurves, we re-parameterise $K$ in terms of the ablation coefficient $\sigma$$=$$c_{h}/c_{d} H^{\ast}$ and the shape factor $A_{0}$$=$$S_{0}\rho^{2/3}_{m}/M^{2/3}_{0}$ of a meteoroid when it enters the atmosphere, to obtain
\begin{linenomath*}
\begin{equation}
\label{eq:Kdef2}
K= \frac{(1 - \mu)\mbox{ }\mbox{$c_{d}$$A_{0}$}\mbox{ } \sigma \rho_{0} h_{0}V^{2}_{0}}{2 M^{1/3}_{0}\rho^{2/3}_{m}\sin{\gamma}}\mbox{.}
\end{equation}
\end{linenomath*}
The product $c_{d}$$A_{0}$ of the initial shape factor and the drag coefficient in Eq.~\ref{eq:Kdef2} is often assumed to fall within the range 1.55--1.8 for a realistically shaped body \citep{GritsevichPopelenskaya2008,Gritsevich.et.al2017,Shober.et.al2022} and equals 1.21 for an initially spherical meteoroid. 

Eqs.~\ref{eq:intensity}, \ref{eq:masslossheight} \& \ref{eq:Kdef2} can now be used to generate meteor lightcurves for small meteoroids of various speeds and masses. Figure~\ref{FIG:earthvenusmeteor} shows such a lightcurve for an initial mass $M_{0}$$=$0.25 g, density $\rho_{m}$=1000 kg $\mbox{m}^{-3}$, speed $V_{0}$=30 km $\mbox{s}^{-1}$ and $\gamma$=$45^{\circ}$. The meteoroid luminous efficiency $\tau$, ablation coefficient $\sigma$ and $\mu$ were fixed at the values $\tau$=1\%, $\sigma$=0.1 $\mbox{s}^{2}$ $\mbox{km}^{-2}$ and $\mu$=0.65 respectively. Luminous intensity is converted to stellar magnitude through the expression \citep{CeplechaRevelle2005,GritsevichKoschny2011}
\begin{linenomath*}
\begin{equation}
\label{eq:magnitude}
m=-2.5 \left(\log_{10}{I} - 3.185\right)
\end{equation}
\end{linenomath*}
with $I$ expressed in W. The resulting meteor reaches $m$$\simeq$$+0^{m}$ at $h_{max,E}$$\simeq$89 km in the Earth's atmosphere; the slightly brighter meteor at Venus has $h_{max,V}$$\simeq$106 km.

\begin{figure}
\centering
\includegraphics[width=0.45\columnwidth]{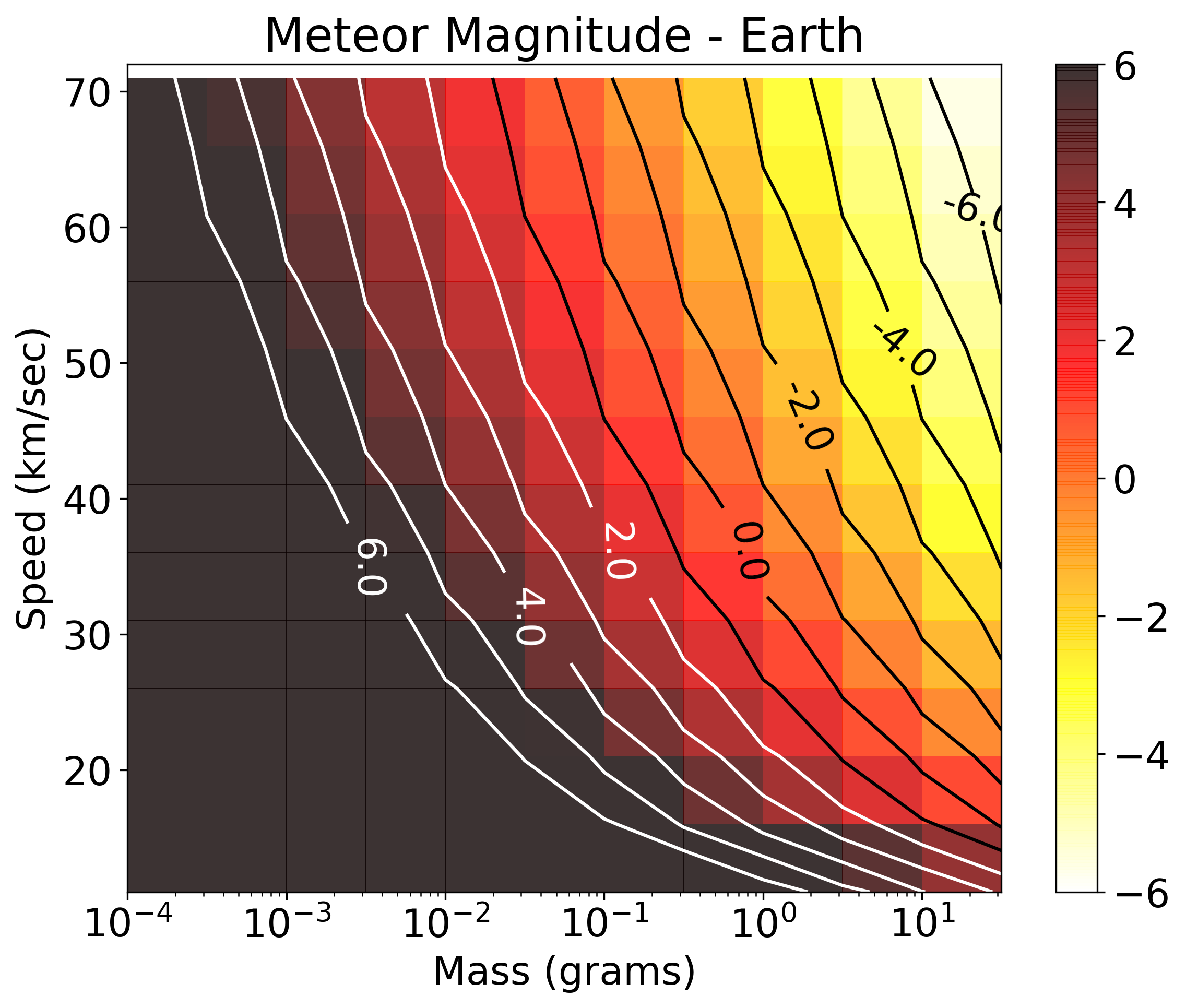}\includegraphics[width=0.45\columnwidth]{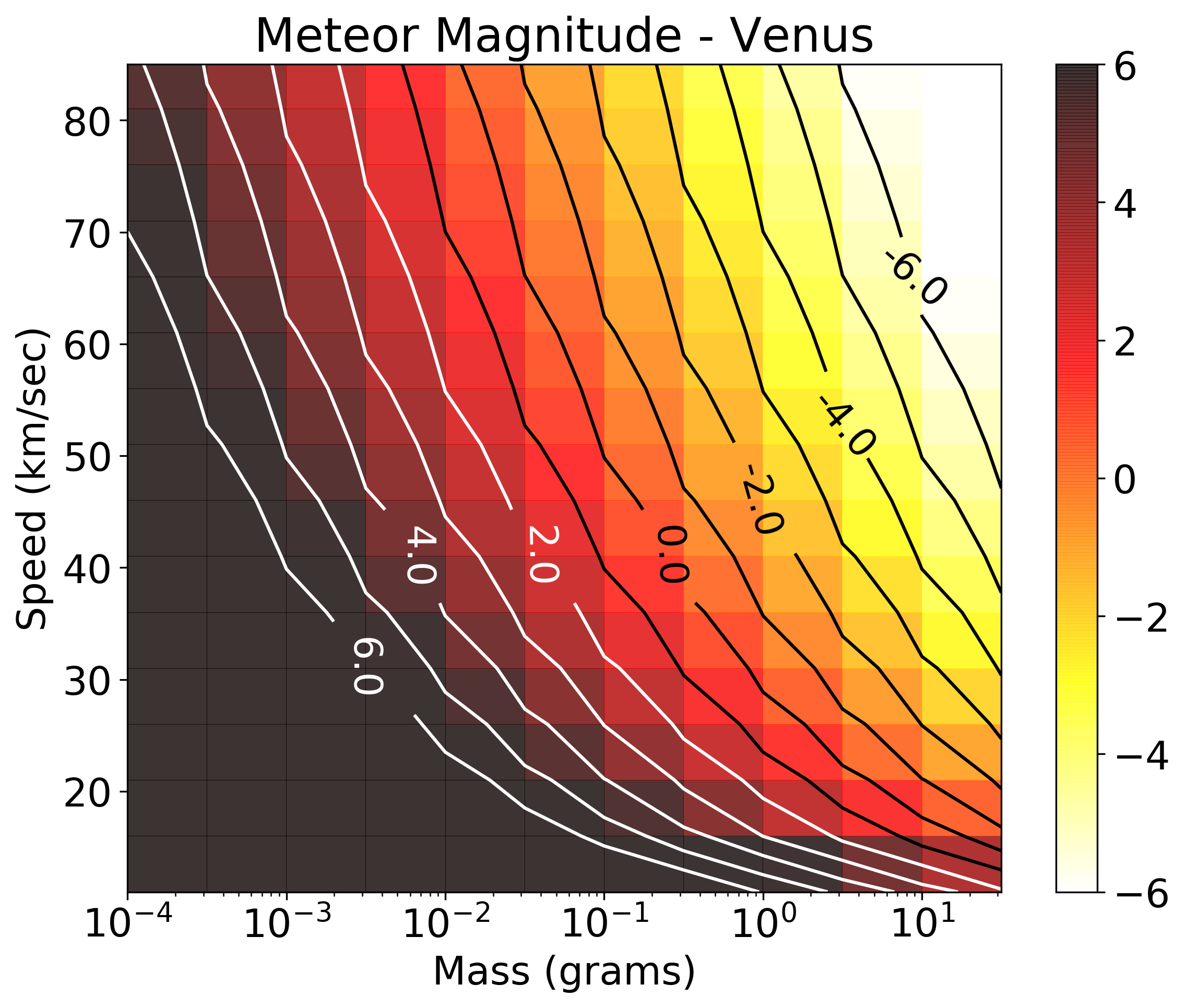}
\includegraphics[width=0.45\columnwidth]{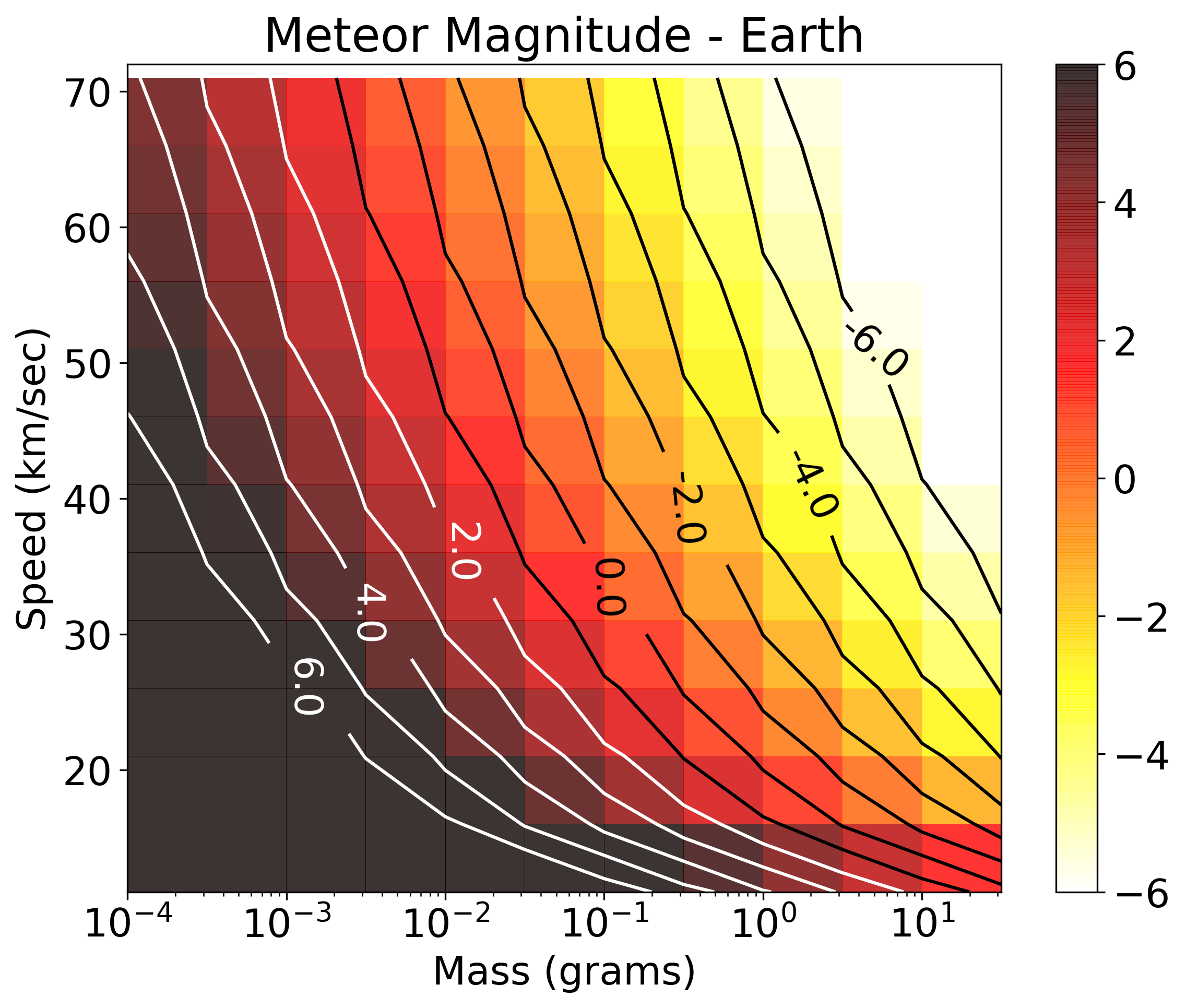}\includegraphics[width=0.45\columnwidth]{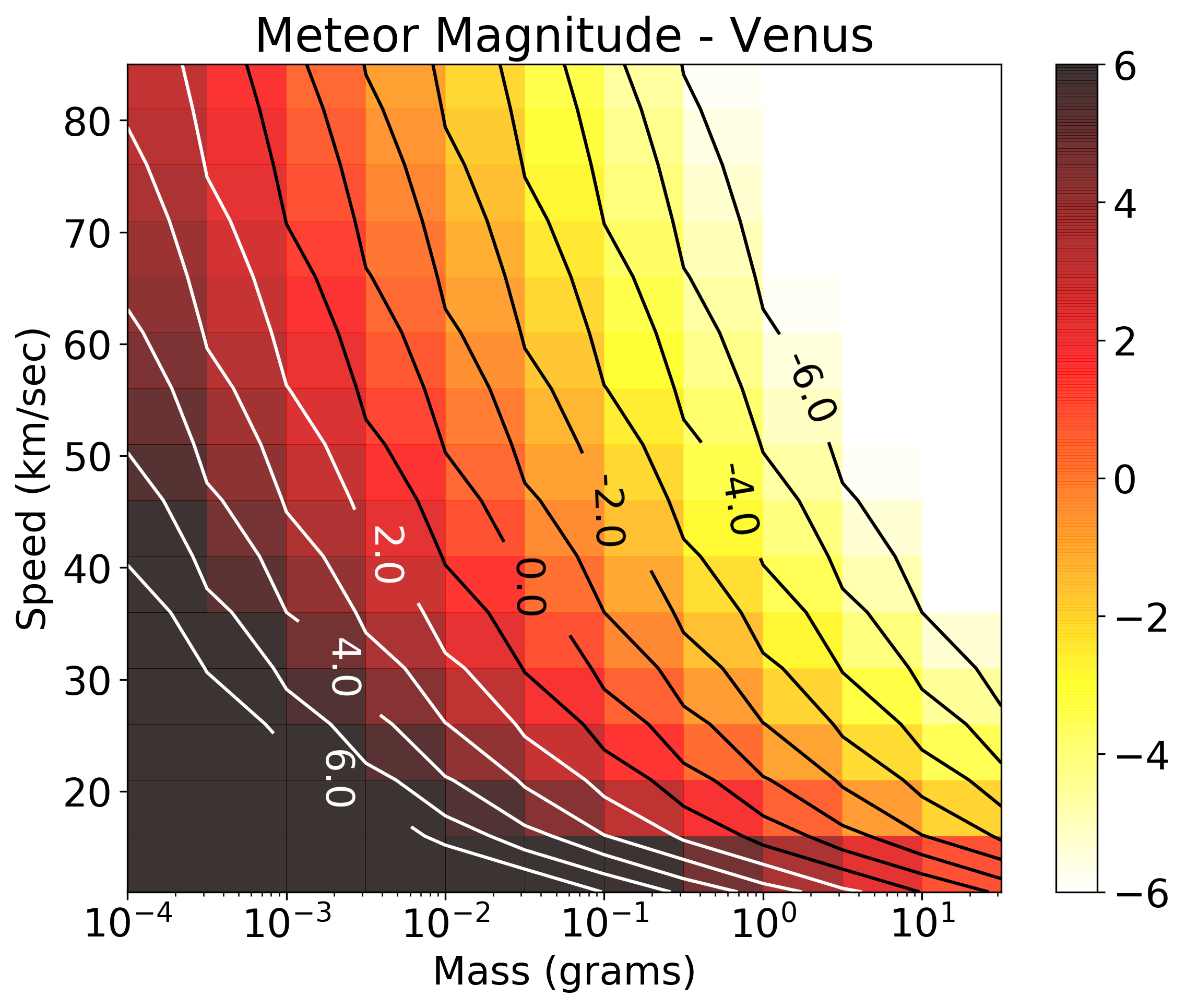}
\caption{Maximum brightness of Earth and Venus meteors for a meteoroid density of 500 kg $\mbox{m}^{-3}$ (top) and 2800 kg $\mbox{m}^{-3}$ (bottom). Speed limits correspond to the planetary escape velocity (lower limit) and a head-on encounter with the planet while in a retrograde heliocentric orbit with perihelion at the planet's distance from the Sun (upper limit). See also Table~\ref{TBL:Planets}.}
\label{FIG:MaxMag}
\end{figure}

It can be shown that, for a given set of meteoroid initial properties the relative maximum intensity of meteors at the two planets depends only on the ratio of atmospheric scale heights 
\begin{linenomath*}
\begin{equation}
\label{eq:relative}
\frac{I_{max,V}}{I_{max,E}} = \frac{h_{0E}}{h_{0V}}
\end{equation}
\end{linenomath*}
or $\Delta m = -0^{\rm m}\hspace{-0.1cm}.63$ (Eq.~\ref{eq:magnitude}) therefore meteoroids of the same mass, density and speed at Venus will always produce brighter but shorter-lived meteors than their terrestrial counterparts. Model brightnesses as functions of mass and speed are shown in Fig.~\ref{FIG:MaxMag} for meteoroid bulk densities of 500 and 2800 kg $\mbox{m}^{-3}$, used here to represent cometary and asteroidal material respectively.
\begin{figure}
\centering
\includegraphics[width=0.45\columnwidth]{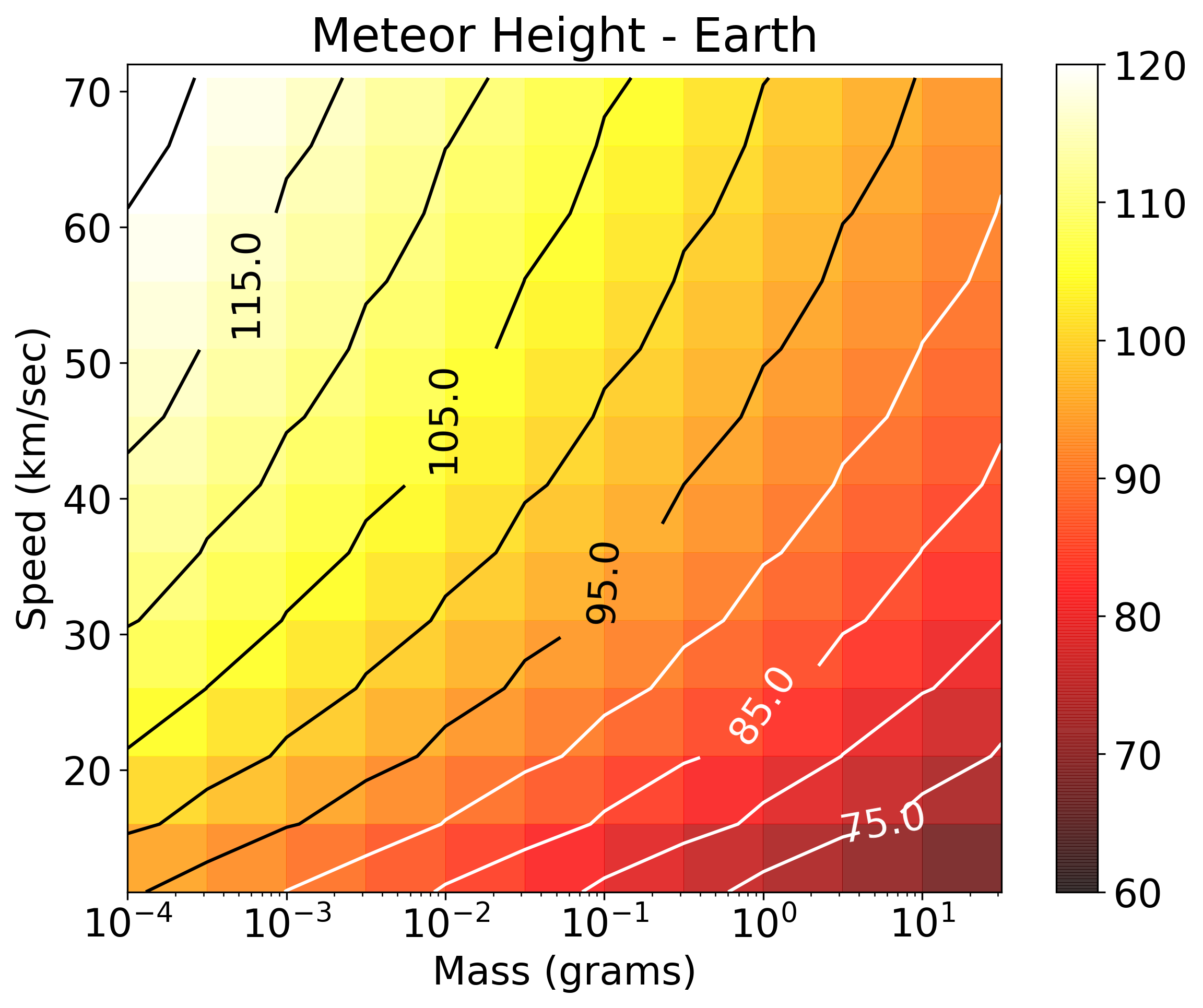}\includegraphics[width=0.45\columnwidth]{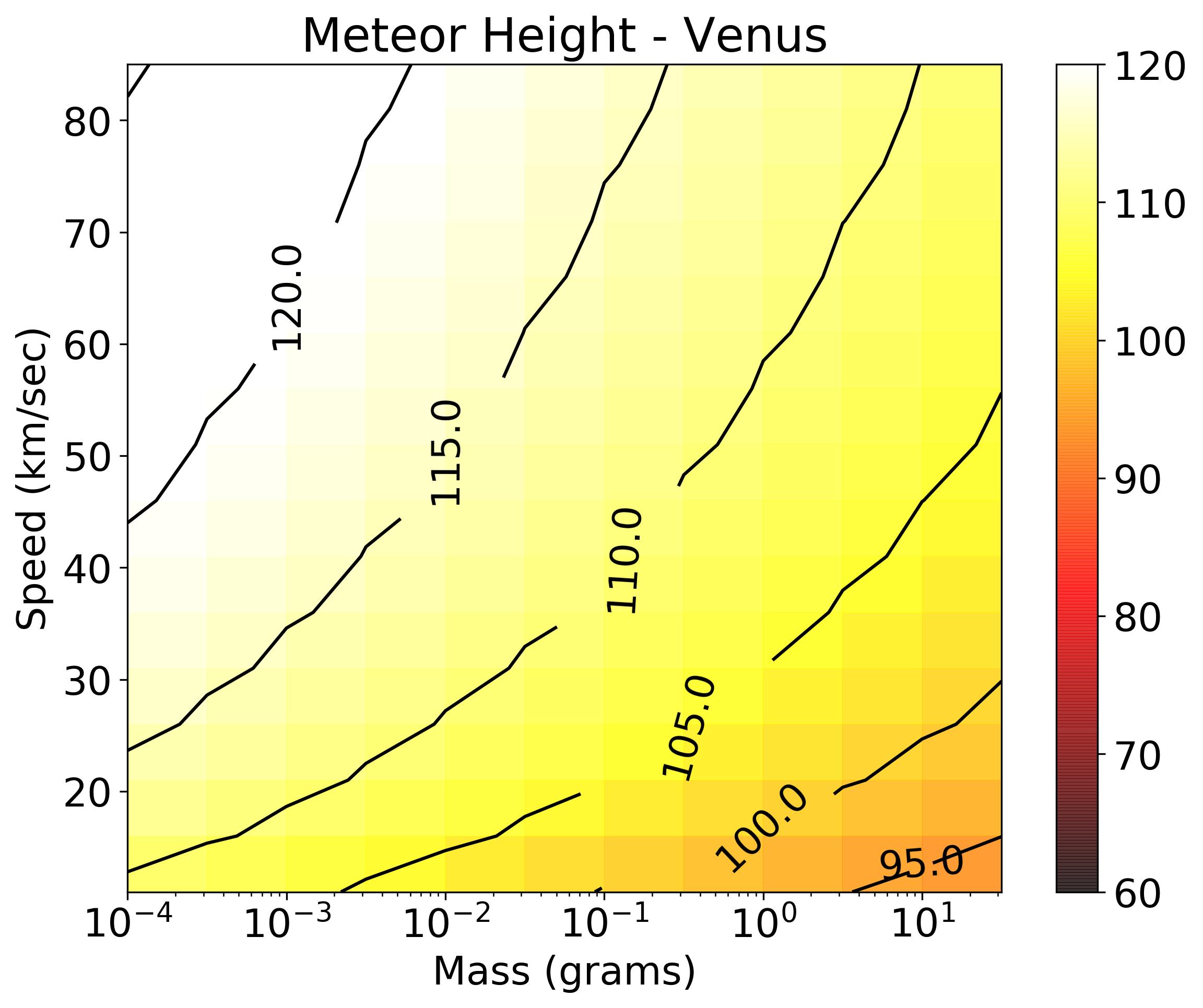}
\includegraphics[width=0.45\columnwidth]{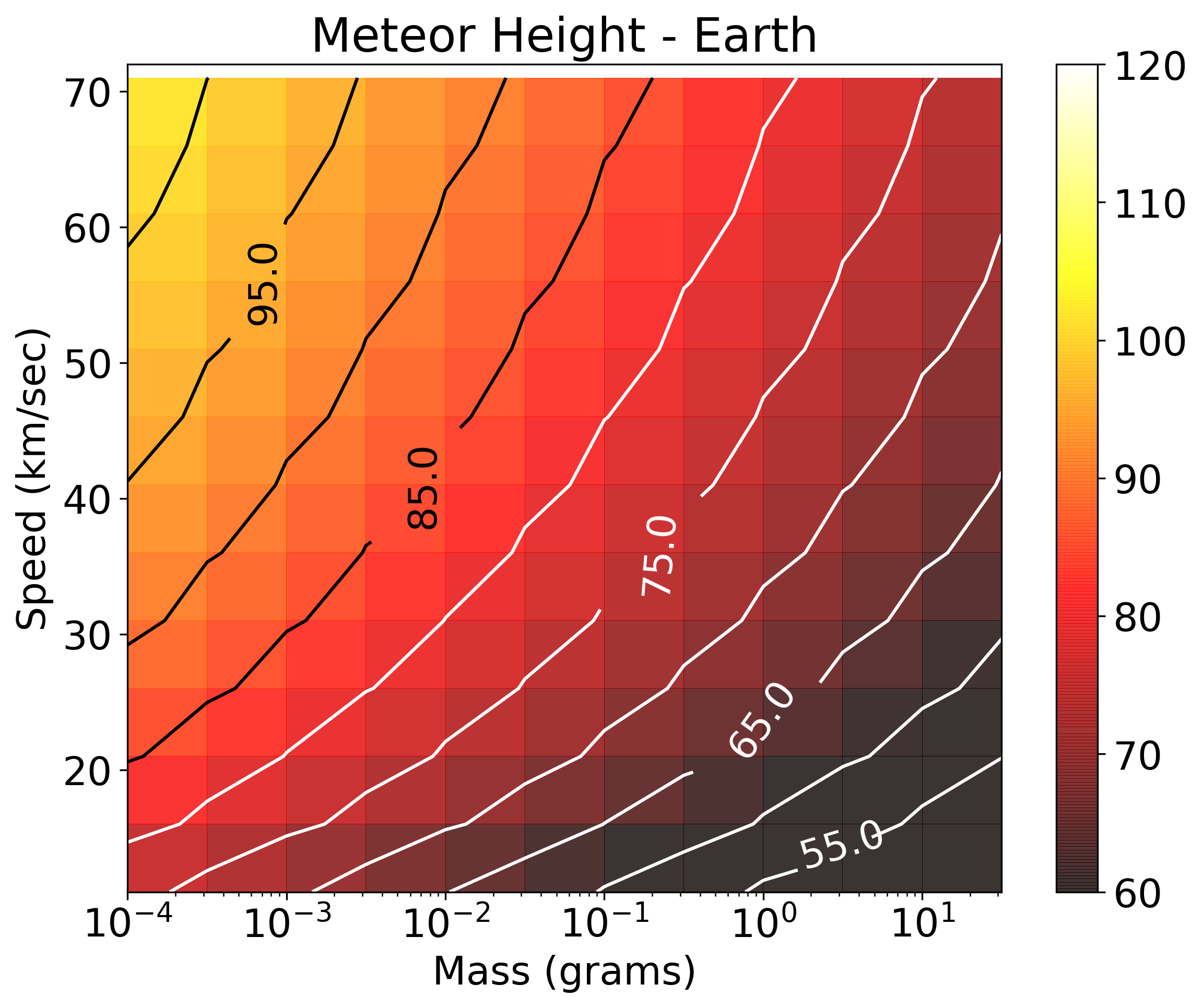}\includegraphics[width=0.45\columnwidth]{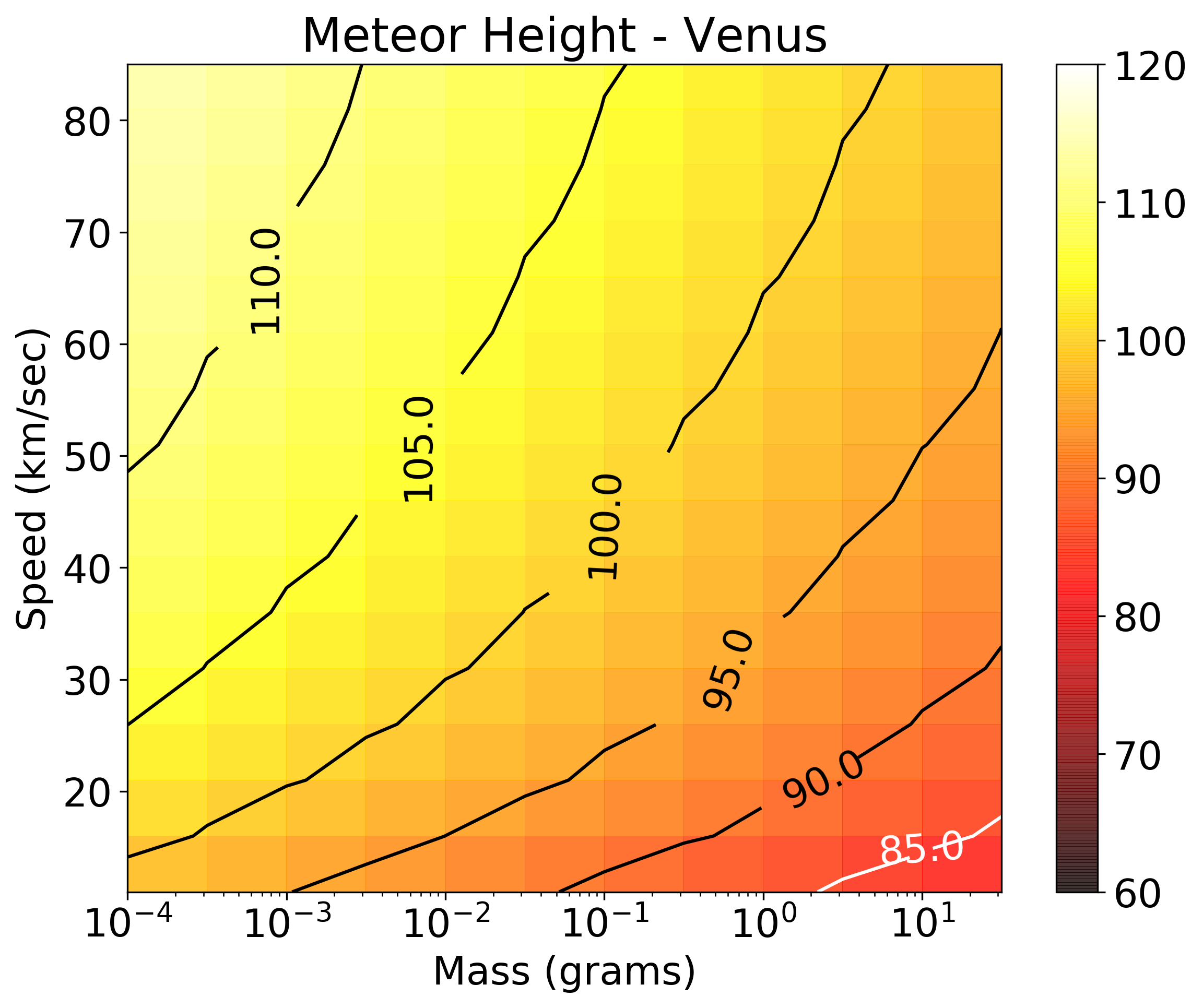}
\caption{Height at maximum brightness for Earth and Venus meteors for meteoroid densities of 500 kg $\mbox{m}^{3}$ (top) and 2800 kg $\mbox{m}^{3}$ (bottom). A brighter colour indicates a higher altitude.}
\label{FIG:HeightMag}
\end{figure}
These values are in good agreement with the ablation simulations of \citet{McAuliffeChristou2006a} that used the actual atmospheric density-height profiles from \citet{Seiff1983}. Those authors reported Earth $-$ Venus differences between +0.4$^{\rm m}$ and +0.7${}^{\rm m}$ for high-density rocky material while the same differences for low-density icy material and masses $<$ $10^{-4}$ g were higher, $>$+1.0$^{\rm m}$.

Similarly, we find that model peak ablation {\it heights} of terrestrial meteors are 75--120 km in the cometary case and 55--100 km in the asteroidal case (Fig.~\ref{FIG:HeightMag}). Theory predicts higher ablation heights at Venus and, for the same mass and speed ranges, we find these to be 95--125 km for cometary meteors and 85--110 km for asteroidal meteors.
These brightness and height values are again in very good agreement with the ablation simulations of \citet{McAuliffeChristou2006a} for meteoroids in the same entry mass and speed ranges.

Actually, in the exponential atmosphere approximation there is a locus of points in ($M_{0}$,$V_{0}$) space where $\Delta h_{max}$$=$$0$, ie the ablating meteoroids reach maximum luminosity at the same common altitude $h_{C}$ in the two atmospheres. This follows from setting $h_{max}$=$h_{V,max}$=$h_{E,max}$ in Eq.~\ref{eq:maxheight}, obtaining the expression
\begin{linenomath*}
\begin{equation}
\label{eq:common_height}
h_{C} =\frac{h_{0E}h_{0V}}{h_{0E}-h_{0V}} \log{\left(\frac{\rho_{0E} h_{0E}}{\rho_{0V}h_{0V}}\right)}
\end{equation}
\end{linenomath*}
that evaluates to $h_{C}$=$125.9$ km using the atmospheric parameter values in this paper. Moreover, we have that $|h_{V} - h_{C}|<|h_{E} - h_{C}|$, in other words Venus meteors reach maximum brightness within a narrower range of altitudes than at Earth. As the Venusian atmosphere is cloud- and haze-free above 90 km altitude \citep{Titov.et.al2018}, meteor surveys from orbiters around Venus appear feasible, a conclusion also reached by \citet{McAuliffeChristou2006a}.

\section{Simulating meteor surveys from orbit}
To assess the performance of a meteor camera onboard a Venus orbiter, we utilize SWARMS (Simulator for Wide Area Recording of Meteors from Space), a {\tt python}-based numerical tool designed to simulate the detection of meteors with space-based optical sensors \citep{Bouquet2013,Bouquet.et.al2014}. The tool incorporates analytical equations developed through a physics-based parameterisation to characterize the variations in mass, height, velocity, and luminosity of a meteor along its atmospheric trajectory \citep{Gritsevich2007,Gritsevich2009,GritsevichKoschny2011,pena2021accurate}. Observations are assumed to take place from a circular orbit at a fixed altitude $H$ above the planetary surface, while the detector itself is parameterised by the field of view (FOV), meteor limiting magnitude ($m_{lim}$), exposure time ($T$) and the minimum number of frames ($\mbox{\it ND}$) required for a valid detection. The meteoroid influx is modelled by sampling user-defined statistical distributions of number influx vs mass, of the bulk density and of the entry speed. The surface of the planet is represented by a mesh of equal-area flat elements with the meteor generation and detection algorithms applied only to those mesh elements within the detector field of view and on the planetary hemisphere below the observer.  

SWARMS regards meteors as detected if they exceed a luminosity threshold $I_{min}$ for at least $\mbox{\it ND}\times T$ sec. To simplify the detection decision, lightcurves are assumed gaussian in shape and an empirical factor $F$ is introduced so that detection is determined from the total luminous energy $E$, if
\begin{linenomath*}
\begin{equation}
\label{eq:detection}
E > I_{min} \times T \times \mbox{\it ND} \times F
\end{equation}
\end{linenomath*}
with $E = \tau_{G} E_{kin}$, $E_{kin}$ being the initial kinetic energy.

In the simulations reported in \citet{Bouquet.et.al2014}, both $\tau_{G}$ and $F$ are determined by applying the ablation theory of \citet{Gritsevich2009} and \citet{GritsevichKoschny2011} to the fireball data of \citet{Halliday.et.al1996}. Specifically, $\tau_{G}$ is defined through the empirical function
\begin{linenomath*}
\begin{equation}
\label{eq:taudef}
\tau_{G} = 0.0051 (V_{0} -10 \mbox{ km}\mbox{ s}^{-1})^{0.87}(100\sigma)^{-1.46}
\end{equation}
\end{linenomath*}
where $V_{0}$ is the atmospheric entry speed and the ablation coefficient $\sigma$ is estimated from the meteoroid density $\rho_{m}$ \citep{RevelleCeplecha2001,Bouquet.et.al2014} as
\begin{linenomath*}
\begin{equation}
\label{eq:sigmadef}
\sigma=\log_{10}{(\rho_{m}-0.25)/4.77}/23.5
\end{equation}
\end{linenomath*}
while the value $F$=$F_{E}$$=$$18.51$ is adopted for the empirical scale factor in Eq~\ref{eq:detection}.
Meteoroid speeds in SWARMS follow the log-normal distribution determined from meteor radar observations \citep{Hunt.et.al2004}:
\begin{linenomath*}
\begin{equation}
\label{eq:speed}
\log_{10}{V_{0}} \sim N(\log_{10}{\mu_{V}},\log^{2}_{10}{\sigma_{V}})
\end{equation}
\end{linenomath*}
where $N(.)$ is the standard notation for the two-parameter gaussian probability density function with parameter values $\mu_{V}$=20 km $\mbox{s}^{-1}$ and $\sigma_{V}$=1.35 km $\mbox{s}^{-1}$. Typical speeds for the respective sporadic populations at the two planets should not differ by more than a few km $\mbox{s}^{-1}$ \citep{CarrilloSanchez.et.al2020}, therefore we adopt the same distribution for the Venus runs. Likewise, we adopt the \citeauthor{Bouquet.et.al2014} distribution for the meteoroid bulk density, so that its values are randomly and uniformly chosen between $1000$ and $4000$ kg $\mbox{m}^{-3}$. While meteoroids with density outside this range are abundant in the observational record e.g.~cometary particles with density $<$1000 kg $\mbox{m}^{-3}$ \citep{Ceplecha.et.al1998}, the adopted distribution represents an acceptable compromise between simulation realism and efficient representation of the principal meteoroid types in the calculation of $\tau_G$ and $\sigma$ \cite[Eq.~\ref{eq:taudef} and \ref{eq:sigmadef} resp., see also Table 5 in][]{RevelleCeplecha2001}.
\begin{figure}
\centering
\includegraphics[width=0.5\columnwidth]{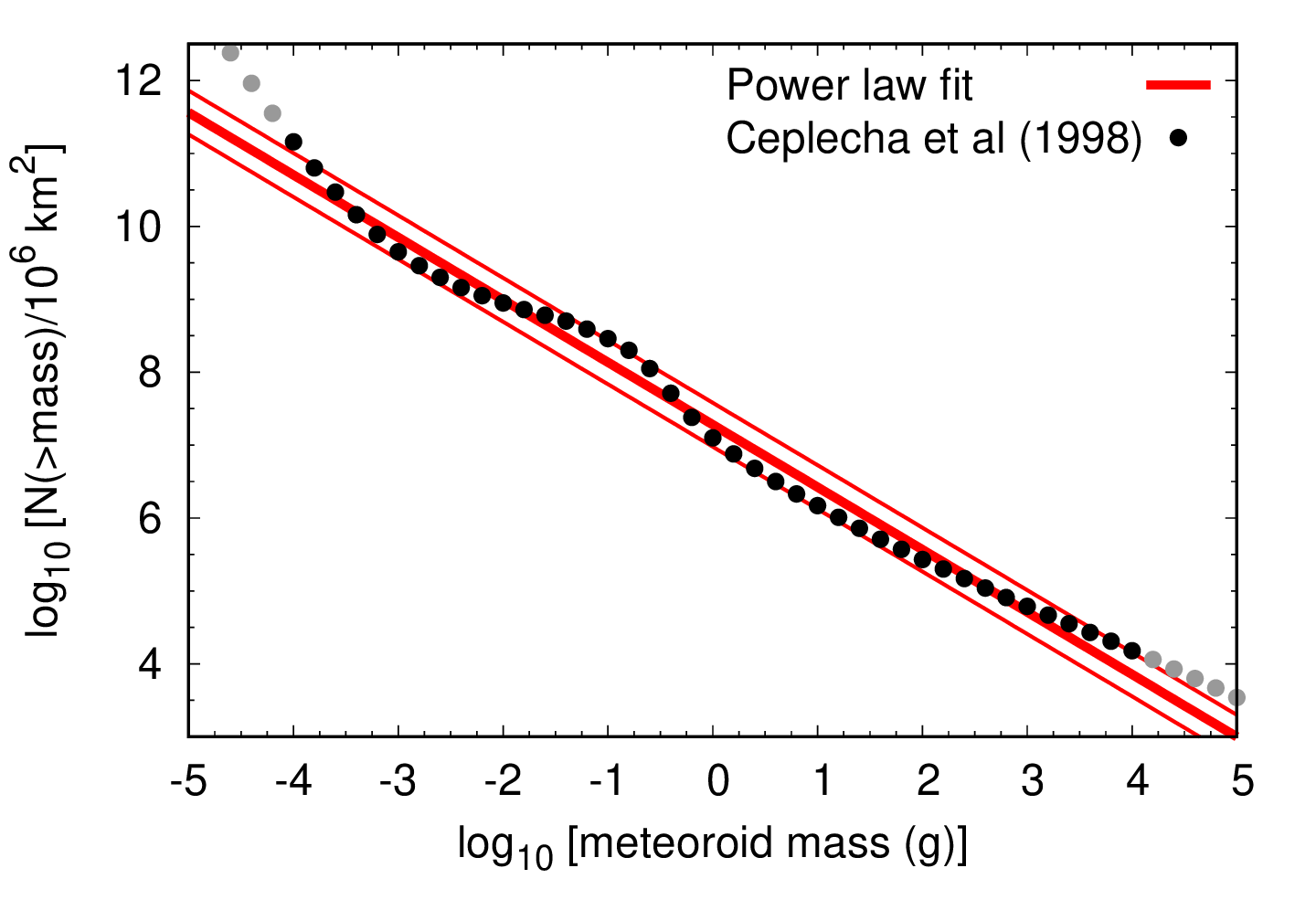}
\includegraphics[width=0.5\columnwidth]{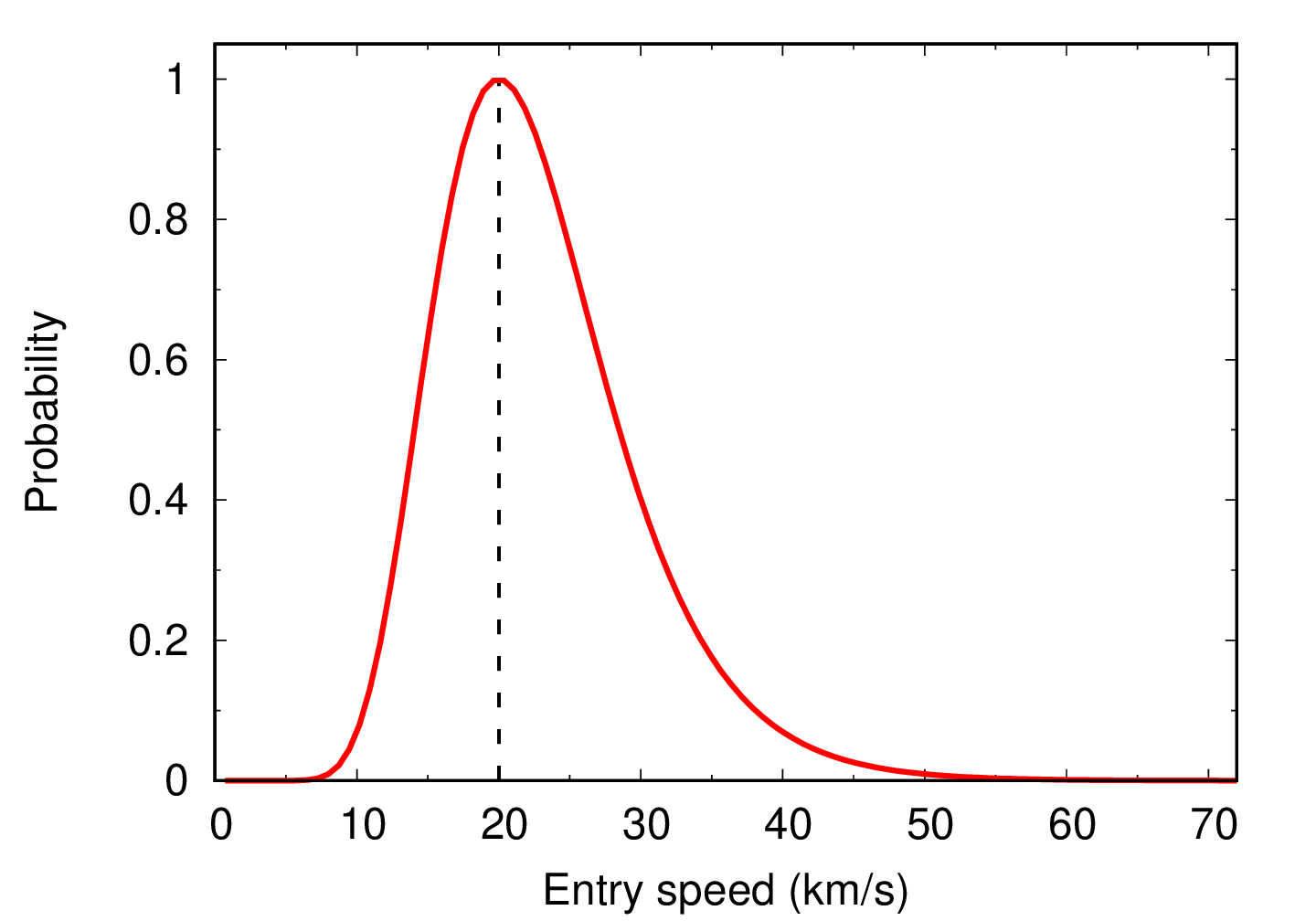}
\caption{Statistical distributions used in SWARMS. Top: Power-law fit (Eq.~\ref{eq:power_law_1e6}) to the flux data in \citet{Ceplecha.et.al1998}. The parallel lines indicate factor-of-two differences from the fit prediction. Grey data points were not considered in the fit. Bottom: Probability density function of the entry speed with a mode <v>=$\mu_{V}$=20 km $\mbox{s}^{-1}$.}
\label{FIG:meteoroid_cfd}
\end{figure}

The SWARMS reference distribution for the meteoroid number influx is that of \citet{Halliday.et.al1996}, obtained from a dataset of fireballs showing noticeable deceleration with absolute magnitudes $m$ between $-4$ and $-13$. These magnitudes are somewhat brighter than those of detectable meteors in our simulations ($m$<$+5$). Moreover, the mass index value of $1.48$ obtained from the fireball data is rather low compared to values closer to $\sim$$2$ reported for these fainter meteors \citep{Ceplecha.et.al1998,Jenniskens.et.al2016}. 
For these reasons, we adopt here a different meteoroid influx reference distribution for our SWARMS simulations than in \citet{Bouquet.et.al2014}.
The new distribution uses the collated data in \citet{Ceplecha.et.al1998} presented as a cumulative distribution of the annual meteoroid number influx to the Earth with masses in the range $10^{-18}$--$10^{18}$ g (cf Table 26). To produce a law suited to our purposes we fitted the function
\begin{linenomath*}
\begin{equation}
\label{eq:power_law}
\log_{10}{N(>M_{0})}=a+b \log_{10}{M_{0}}
\end{equation}
\end{linenomath*}
to these data within the mass interval $10^{-4}$--$10^{4}$ g, yielding $a$=7.230$\pm$0.033 $\log{N(> 1 g)/\mbox{Earth}/\mbox{yr}}$ and $b$=$-$0.856$\pm$0.014. Masses outside this range were not considered in the fit as they do not contribute to meteor detections in the simulations. By further assuming a fixed meteor altitude $h$=$90$ km at Earth we obtain the following expression for the number influx per yr and per $10^{6}$ $\mbox{km}^{2}$ of atmospheric surface area:
\begin{linenomath*}
\begin{equation}
\log_{10}{N(>M_{0})}=4.559-0.856 \log_{10}{M_{0}}
\label{eq:power_law_1e6}
\end{equation}
\end{linenomath*}
This relationship reproduces the \citeauthor{Ceplecha.et.al1998} data to within a factor of two (Fig.~\ref{FIG:meteoroid_cfd}), with the worst agreement observed for $M_{0}$$\simeq$$10^{-4}$ g (equivalent meteoroid diameter of 0.6 mm for a 1000 kg $\mbox{m}^{-3}$ bulk density). 

We then apply SWARMS to either Earth or Venus orbiters by adopting the appropriate set of parameter values from Table~\ref{TBL:Planets} combined with the stated influx, speed and density distributions for the meteoroids. 

Furthermore, we utilise the constant-scale-height fits to the atmospheric profiles shown in Fig.~\ref{FIG:earthvenusprofiles} and divide the scale factor $F_{E}$ in Eq~\ref{eq:detection} by the ratio $h_{0E}/h_{0V}$ to obtain the value $F_{V}$$=$$10.34$ to be used for determining whether a meteor is detectable or not from an orbiter at Venus. No atmospheric extinction correction is applied to the apparent brightness of simulated meteors. We expect this to be neglibigle for typical meteor heights at Earth and Venus (Fig~\ref{FIG:HeightMag}) when observation takes place from above the atmosphere as is the case here. The simulated survey duration in each of our SWARMS runs is varied from a few hours to several weeks, as appropriate to accumulate a sufficiently high number of detections - at least several hundreds - to ensure a statistically robust outcome. This number is then divided by the survey duration to obtain the hourly detection rate for that run. Finally, we do not consider the effects of spatial or temporal variations in the terrestrial and venusian sporadic rate. Although such variations likely exist at both planets \citep{CampbellBrown2007,Janches.et.al2020}, it is not clear to us that they are significant enough to alter the outcome of this study.

\begin{table}
\centering
\caption{Planetary characteristics used in the simulations.}\label{TBL:Planets}
\begin{tabular}{lrr}
\toprule
Property & Earth & Venus \\ \hline \noalign{\smallskip}
Mass ($\times$$10^{24}$ kg) & 5.97 & 4.87  \\
Radius (km) & 6378 &  6051 \\
Meteor altitude (km) & 90 &  110 \\
Escape speed (km $\mbox{s}^{-1}$) &  11.10 & 10.27 \\
Maximum pre-atmospheric speed (km $\mbox{s}^{-1}$)  &   72.8 & 85.2 \\
\bottomrule
\end{tabular}
\end{table}
\begin{table}
\centering
\caption{Characteristics of camera systems in this study.}\label{TBL:Cameras}
\begin{tabular}{crr}
\toprule
Property & SPOSH & Mini-EUSO \\ \hline \noalign{\smallskip}
Field of view (${}^{\circ}$) & 120 &  40 \\
Limiting magnitude &  +6 & +5 \\
Exposure time (s) & 0.06 & 0.04\\
Angular speed (${}^{\circ}\mbox{ s}^{-1}$)& 5 & 5 \\
\bottomrule
\end{tabular}
\end{table}
\section{Scientific use cases: SPOSH and Mini-EUSO}
Apart from the model meteoroid population and the observation circumstances, the degree of realism in our simulations depends on selecting detector characteristics appropriate for the task. A wide field of view, high sensitivity, and a rapid frame rate are all essential requirements for the efficient optical detection of meteors, yet instrumentation with flight heritage seldom features these in combination. 

Recent interest in detecting fast optical transients in the Earth's atmosphere from space has yielded two facility concepts suitable for the task: the Smart Panoramic Optical Sensor Head \cite[SPOSH;][]{Oberst.et.al2011,Christou.et.al2012} and the Joint Exploratory Missions for an Extreme Universe Space Observatory \cite[JEM-EUSO;][]{Haungs.et.al2015}. Breadboard versions of SPOSH were successfully field-tested in a two-station configuration, completing a multi-year survey of the Perseid meteor shower \citep{Margonis.et.al2019} while the JEM-EUSO study evolved into the Mini-EUSO demonstrator that has been systematically recording meteors from the International Space Station since 2019 \citep{Bacholle.et.al2021,Coleman.et.al2023}.

The principal characteristics of SPOSH and Mini-EUSO are listed in Table~\ref{TBL:Cameras}. 
\begin{figure}
\centering
\includegraphics[width=0.5\columnwidth]{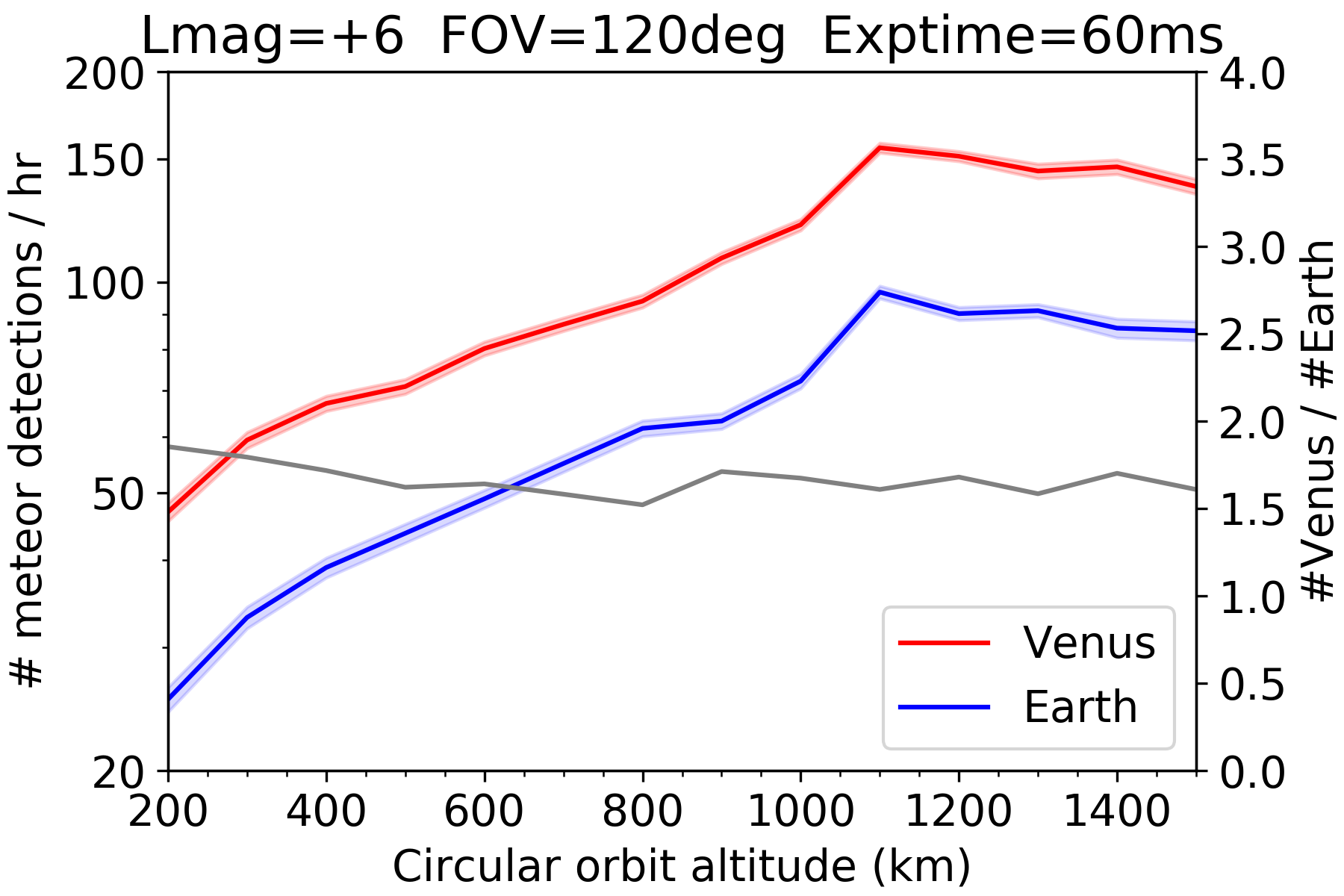}
\includegraphics[width=0.5\columnwidth]{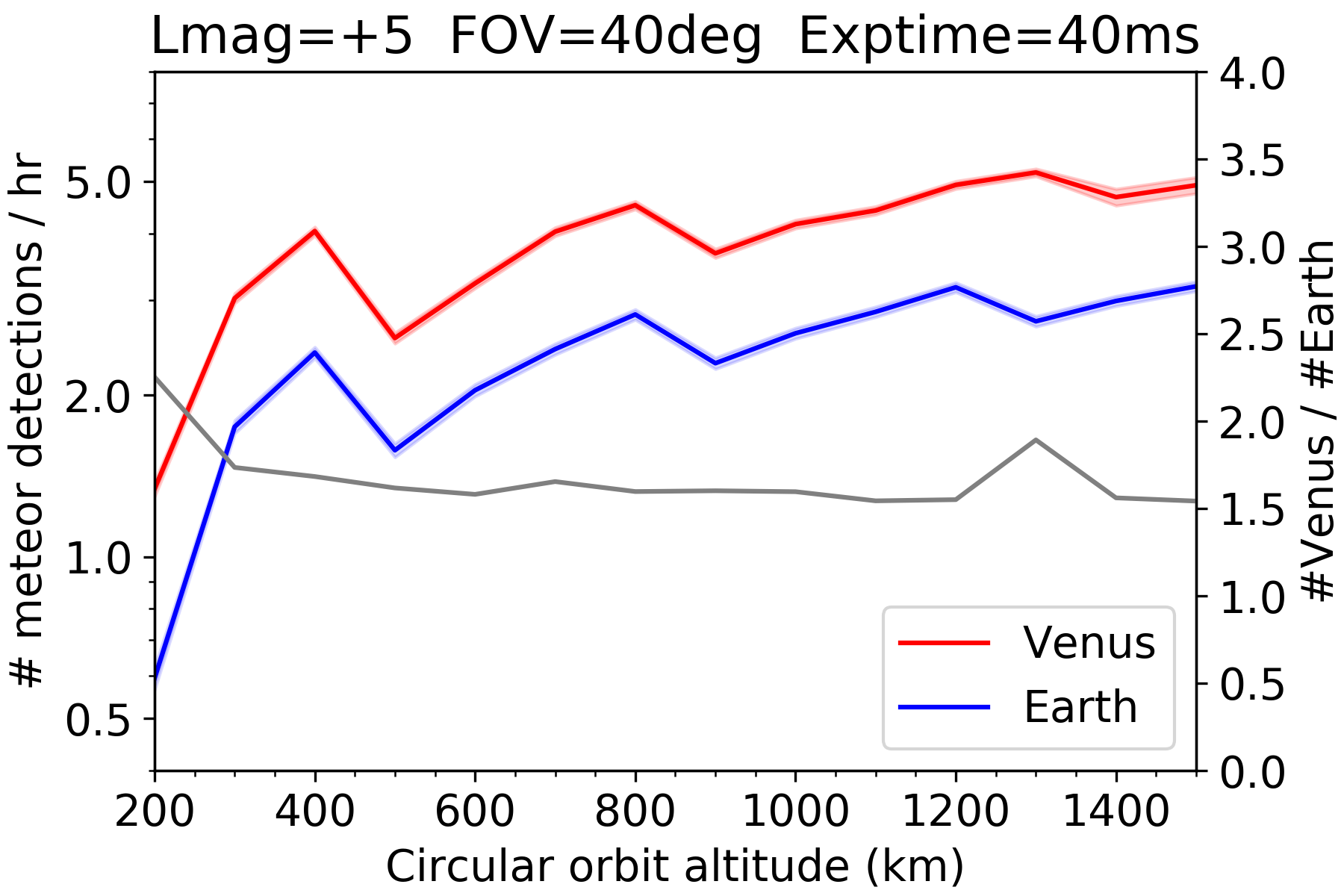}
\caption{Hourly rate of meteor detections at Earth and Venus for detectors with similar characteristics to SPOSH (top) and Mini-EUSO (bottom). The gray line represents the ratio of the number of meteors detected at the two planets.}
\label{FIG:n_vs_altitude}
\end{figure}
The SPOSH camera, also adopted by \citeauthor{Bouquet.et.al2014} in their study, was specifically designed to detect meteors from orbital altitudes \citep{Oberst.et.al2011} while Mini-EUSO was developed as a proof-of-concept optical transient detector on the International Space Station under the JEM-EUSO project  \citep{Coleman.et.al2023}. Operating in the ultraviolet part of the electromagnetic spectrum (290--430 nm) and designed to detect meteors of apparent magnitude +5 or brighter under dark background conditions \citep{Abdellaoui.et.al2017}, Mini-EUSO has been operating since 2019 mounted on the ISS {\it Zvezda} module at an altitude of 400 km above the Earth's surface \citep{Bacholle.et.al2021}. The actual flight model features a slightly larger FOV than the original design, however for our simulations we adopted the design value of $40^{\circ}$. 
\begin{figure}
\centering
\includegraphics[width=0.48\columnwidth]{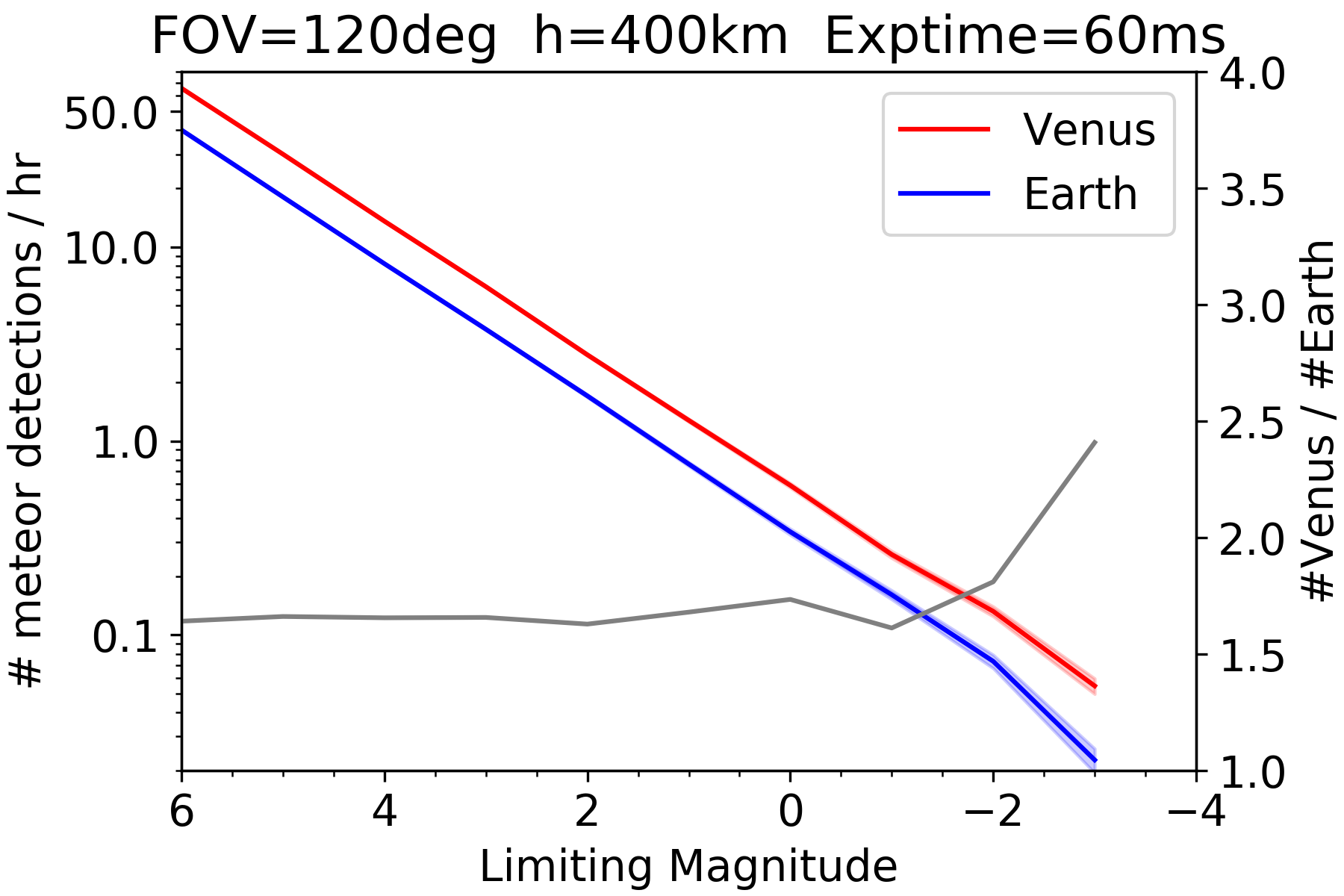}\hspace{2mm}\includegraphics[width=0.48\columnwidth]{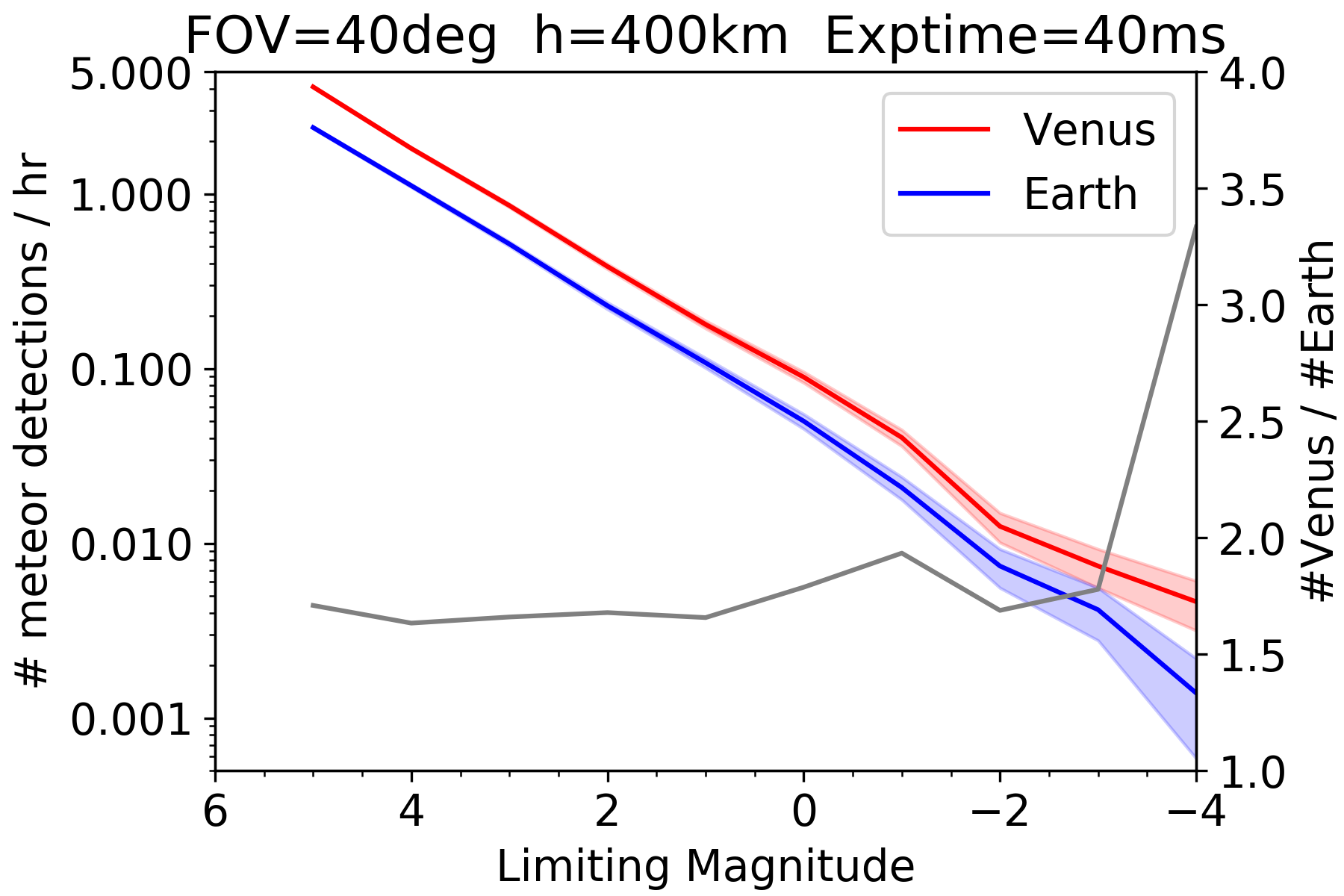}
\includegraphics[width=0.49\columnwidth]{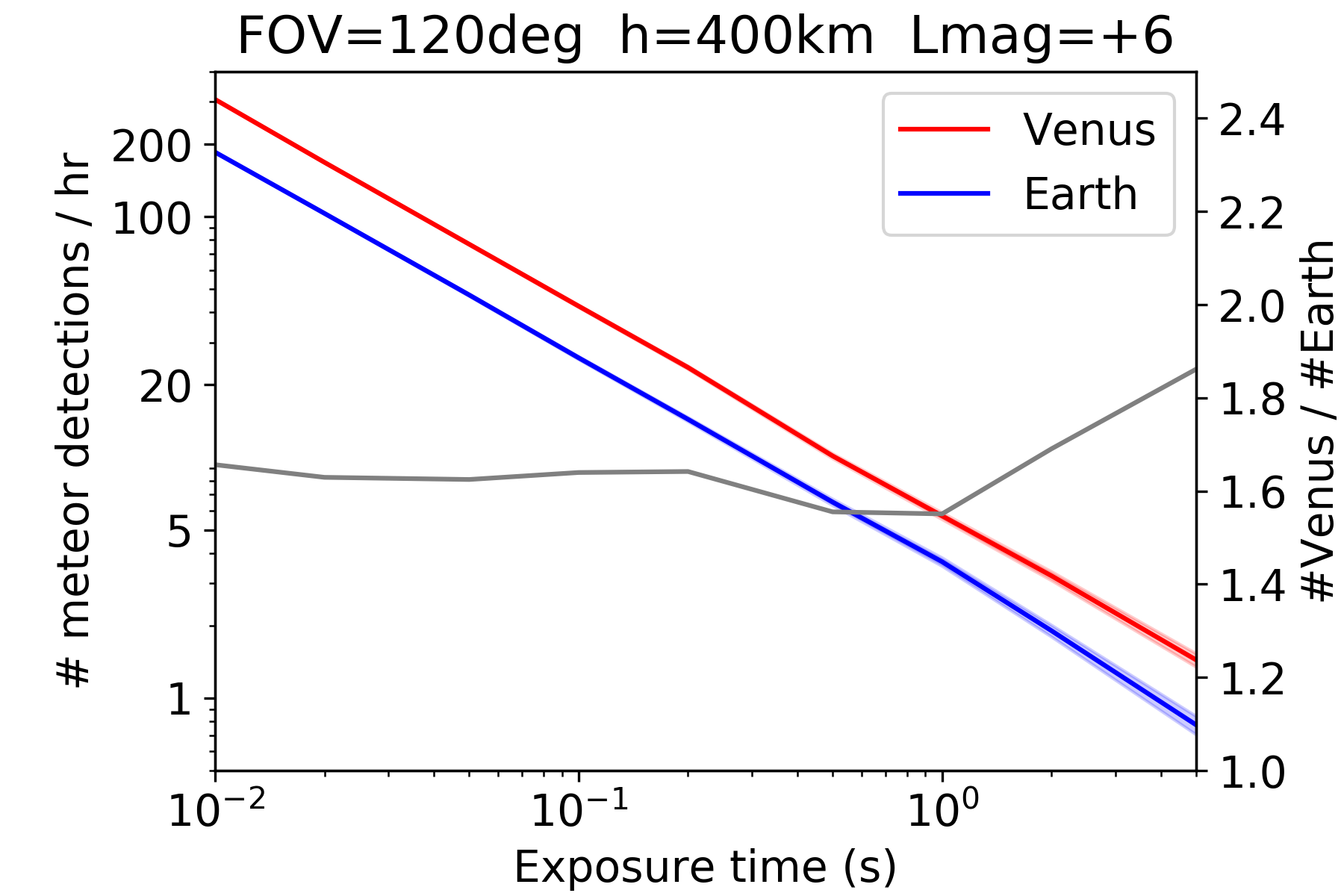}
\hspace{1mm}\includegraphics[width=0.49\columnwidth]{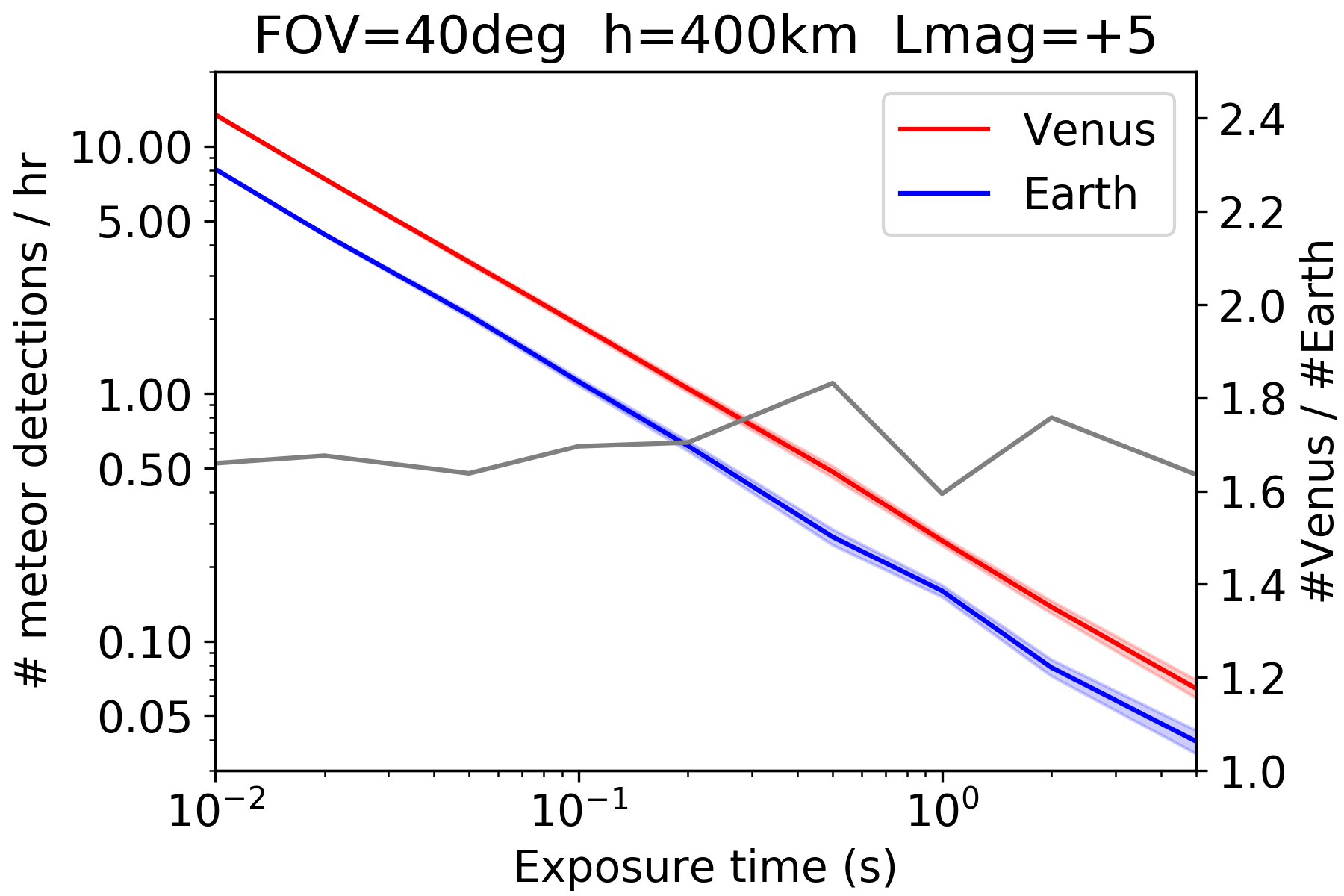}
\includegraphics[width=0.47\columnwidth]{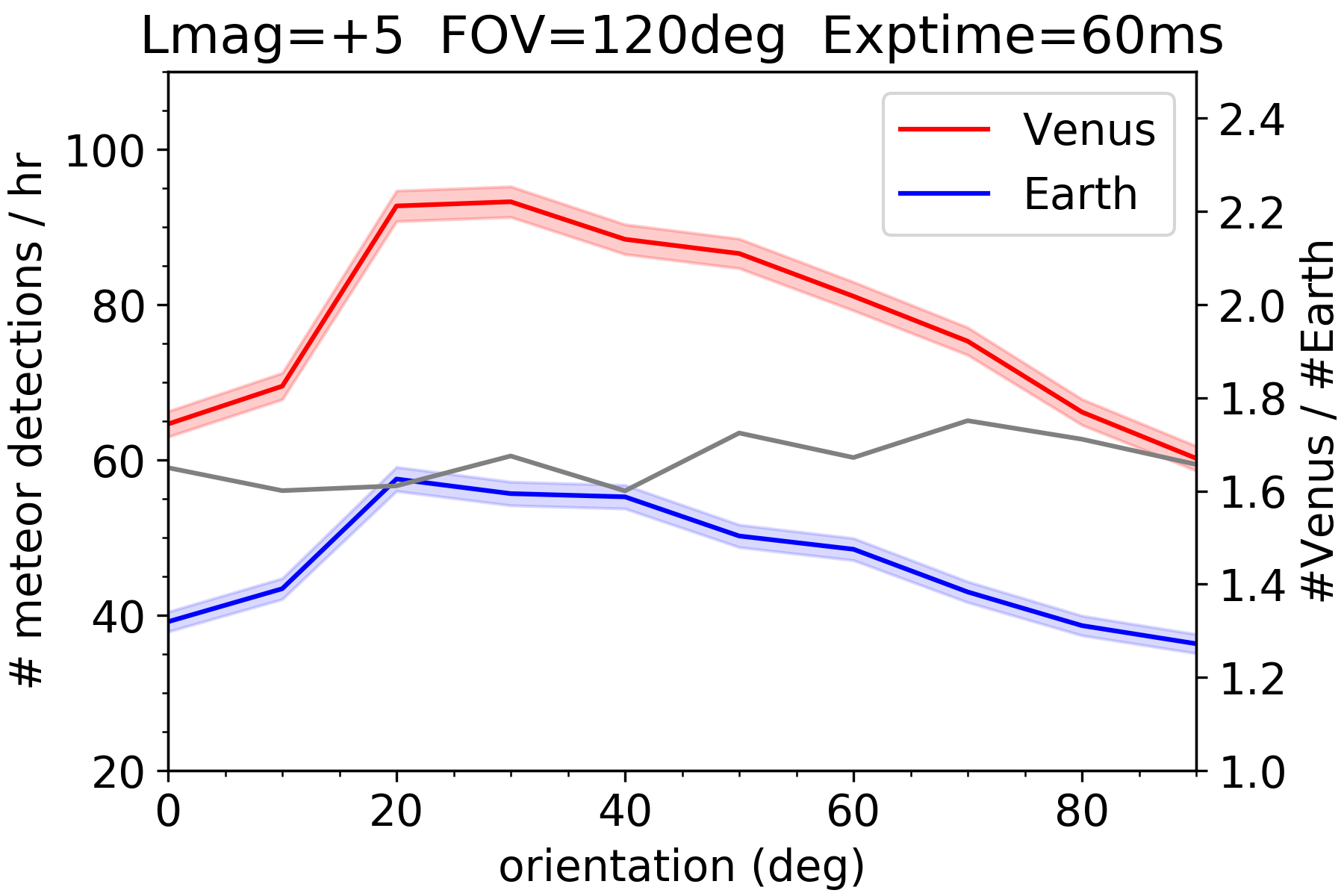}\hspace{3mm}\includegraphics[width=0.47\columnwidth]{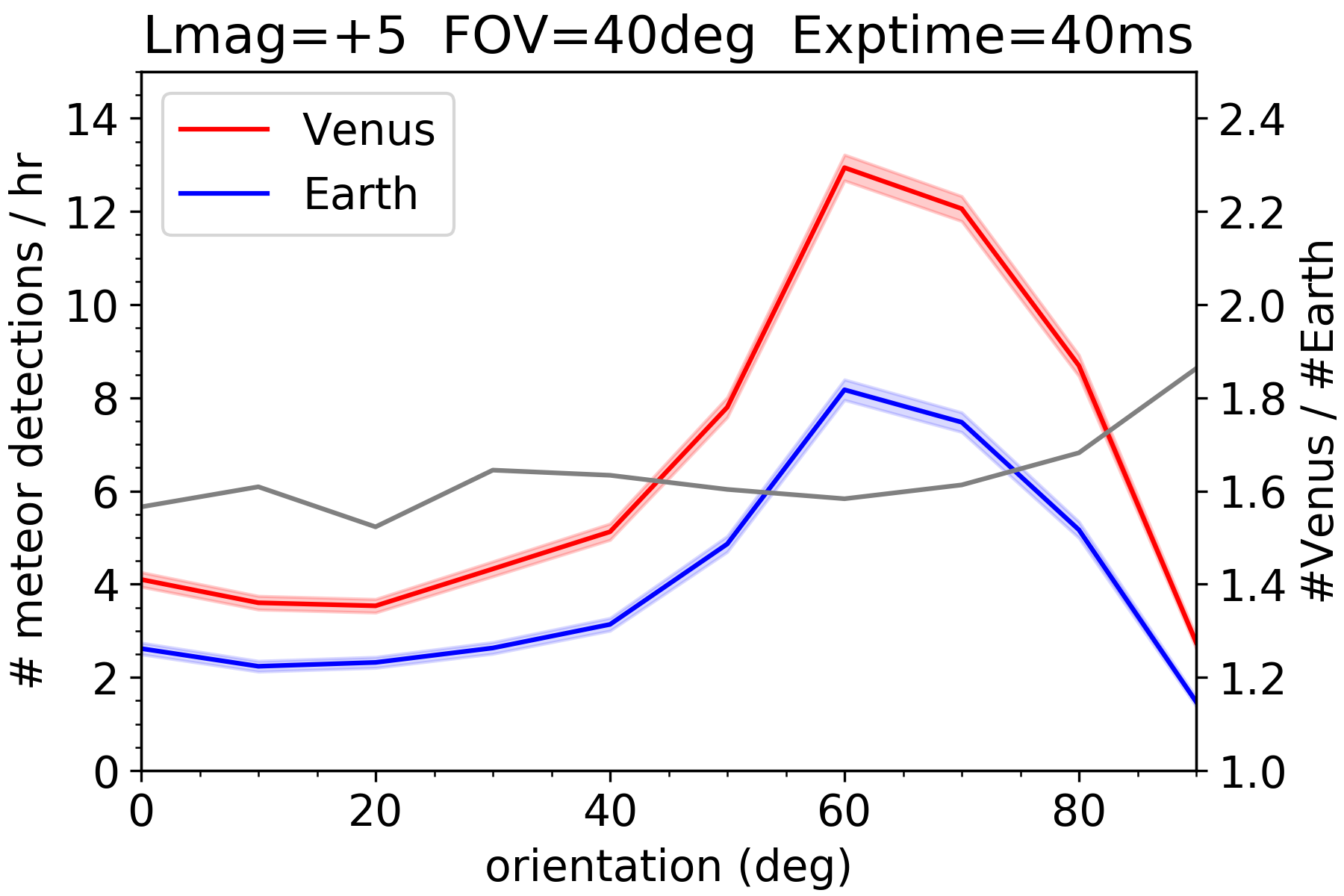}
\caption{Hourly rate of meteor detections at Earth and Venus as a function of camera limiting magnitude (top), exposure time (middle) and orientation (bottom) for SPOSH (left column) and for Mini-EUSO (right column).}
\label{FIG:n_vs_sensitivity}
\end{figure}
Figure \ref{FIG:n_vs_altitude} shows the simulation output for both use cases and for a range of orbital altitudes at Earth and Venus. Narrow shaded bands that accompany each line represent 1-$\sigma$ Poisson counting uncertainties from the simulation output. Consistent with the analysis in Section 2, more meteors are generally detected at Venus than at Earth. The detection rate ratio (grey line) takes values in the range 1.5--2.0 for each camera and is fairly independent of altitude. The primary difference between the use cases is a higher detection rate for SPOSH (25--150 $\mbox{hr}^{-1}$) compared to Mini-EUSO (0.6--5 $\mbox{hr}^{-1}$) owing to the much wider field of view of the former. The character of the altitude dependence is also different; for SPOSH the detection rate gradually increases until $h$$\sim$$1100$ km and remains approximately constant thereafter, while the Mini-EUSO detection rate is roughly constant for $h$$\gtrsim$300 km following a rapid increase. 

The initial increase is likely due to the larger atmospheric surface area surveyed, where this effect dominates over the loss of fainter meteors due to the increasing distance. With a further increase in altitude for the wider-field SPOSH, the planet eventually fills the field of view, leading to the change in slope observed at 1100 km.

\section{Sensitivity analysis}
\subsection{Detector characteristics}
A separate batch of simulations was run to study how the detection rate varies with camera characteristics. We focused on three parameters in particular: limiting magnitude, frame exposure time and angle between the boresight and the nadir direction. This quantity is the tilt angle in \citet{Bouquet.et.al2014} but here we refer to it as the {\it orientation} angle. For the simulations in this and the next subsection the observation orbit altitude was fixed at $H$$=$400 km.

We again find (Fig.~\ref{FIG:n_vs_sensitivity}) that the detection rate ratio is unaffected by the changes, varying between 1.5 and 2.5. The absolute detection rate at either planet is a log-linear function of the sensitivity and of the exposure time (top and middle panels respectively), decreasing by an order of magnitude for every $\sim$3-magnitude reduction in camera sensitivity or an order-of-magnitude increase in exposure time length. The variation with orientation angle (bottom panels) is more modest by contrast. Our SPOSH simulations show variation within a factor of 1.5$\times$ peaking at an orientation angle of $\sim$$30^{\circ}$ while the Mini-EUSO runs show a more pronounced, 3$\times$ variation peaking at $\sim$$60^{\circ}$, similar to that noted in \citet{Bouquet.et.al2014} for a $60^{\circ}$ FOV. These extrema correspond to the maximum atmospheric surface area under observation, again highlighting the relative importance of area coverage over range in attempting to detect meteors from orbit.

\subsection{Meteoroid population parameters}
The model meteoroid population in the simulations is parameterised by the mass index $s$ of the power law that describes the meteoroid flux and by the parameters of the atmospheric entry speed distribution, particularly the mode <v>. Both these parameters are unconstrained by direct observations at Venus and it is of interest to determine how varying either $s$ or <v> affects the outcome. 

\begin{figure}
\centering
\includegraphics[width=0.5\columnwidth]{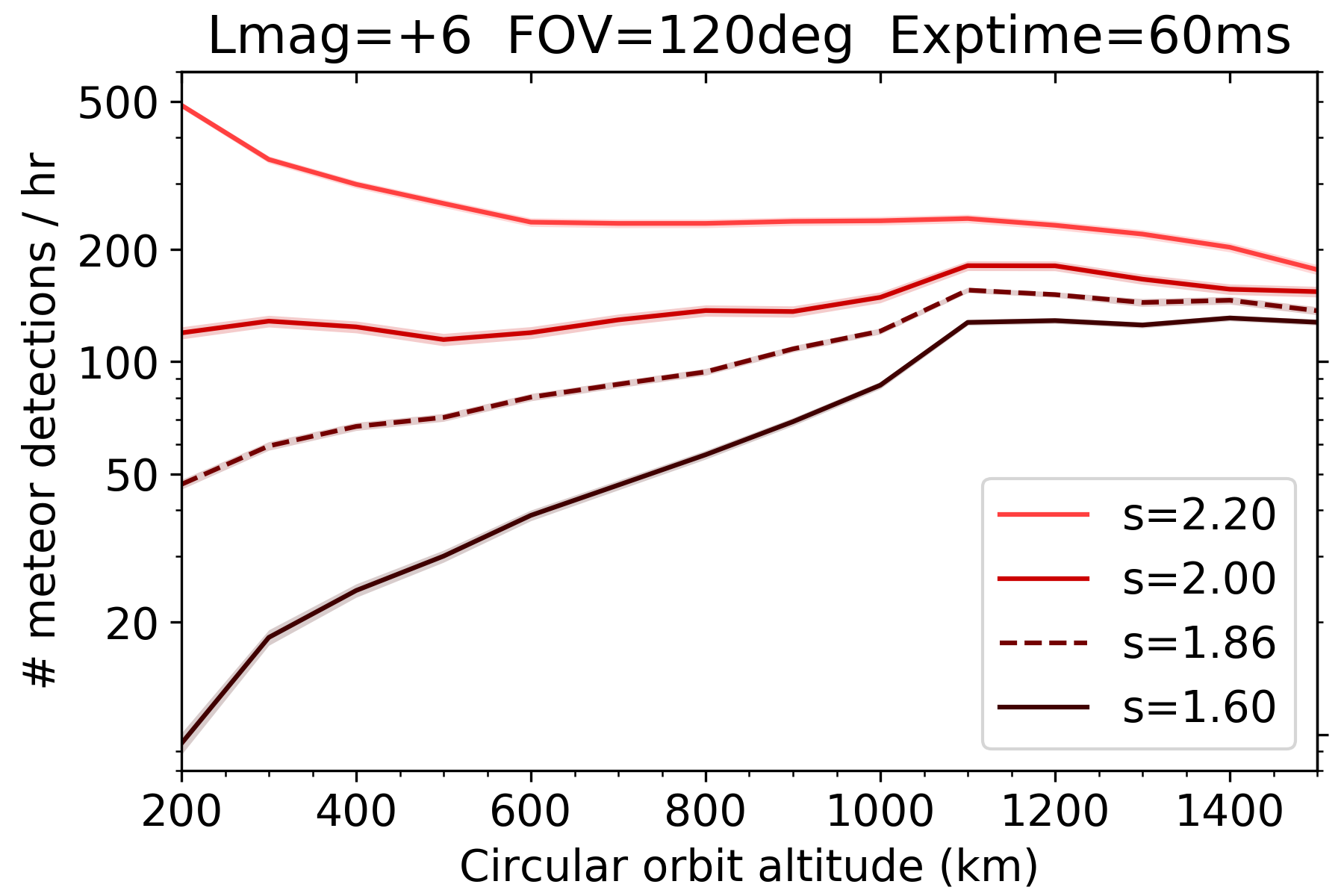}
\includegraphics[width=0.5\columnwidth]{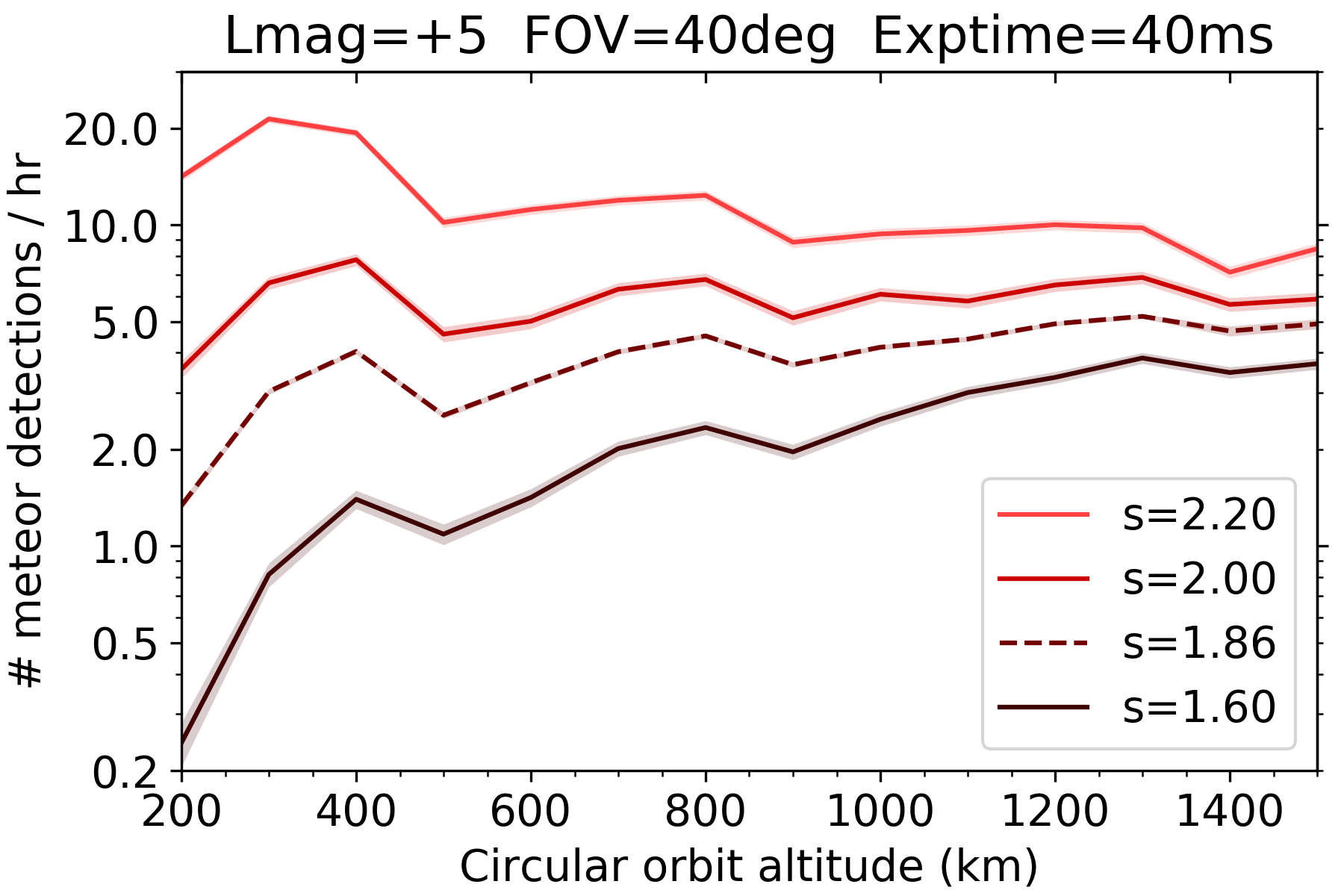}
\caption{Hourly rate of meteor detections at Venus for SPOSH (top) and Mini-EUSO (bottom) for different values of the meteoroid mass index $s$ where a brighter colour indicates a steeper mass distribution.}
\label{FIG:n_vs_altitude_vs_s}
\end{figure}

Figure~\ref{FIG:n_vs_altitude_vs_s} shows the change in the detection rate with altitude for different values of $s$ with the result for the reference value (dashed line) included for comparison. We find that the relative number of detections at lower vs higher altitudes increases with increasing $s$ with the curves flattening out at $s$=2. This flat distribution ($<2\times$ variation at all altitudes) shows that, in this case, the gain in area coverage almost exactly compensates for the loss of the fainter meteors. This feature also appears in our results for Earth (not shown here) and may be a general characteristic of the observation of meteors from space-based platforms. It also means that observational characterisation of meteoroid populations with low $s$ such as meteor showers would benefit from the wider area coverage available at higher altitudes (but see below), or by pointing the camera near the planetary limb (Fig.~\ref{FIG:n_vs_sensitivity}) as in the 1997 Leonid observations \citep{Jenniskens.et.al2000}.
\begin{figure}
\centering
\includegraphics[width=0.5\columnwidth]{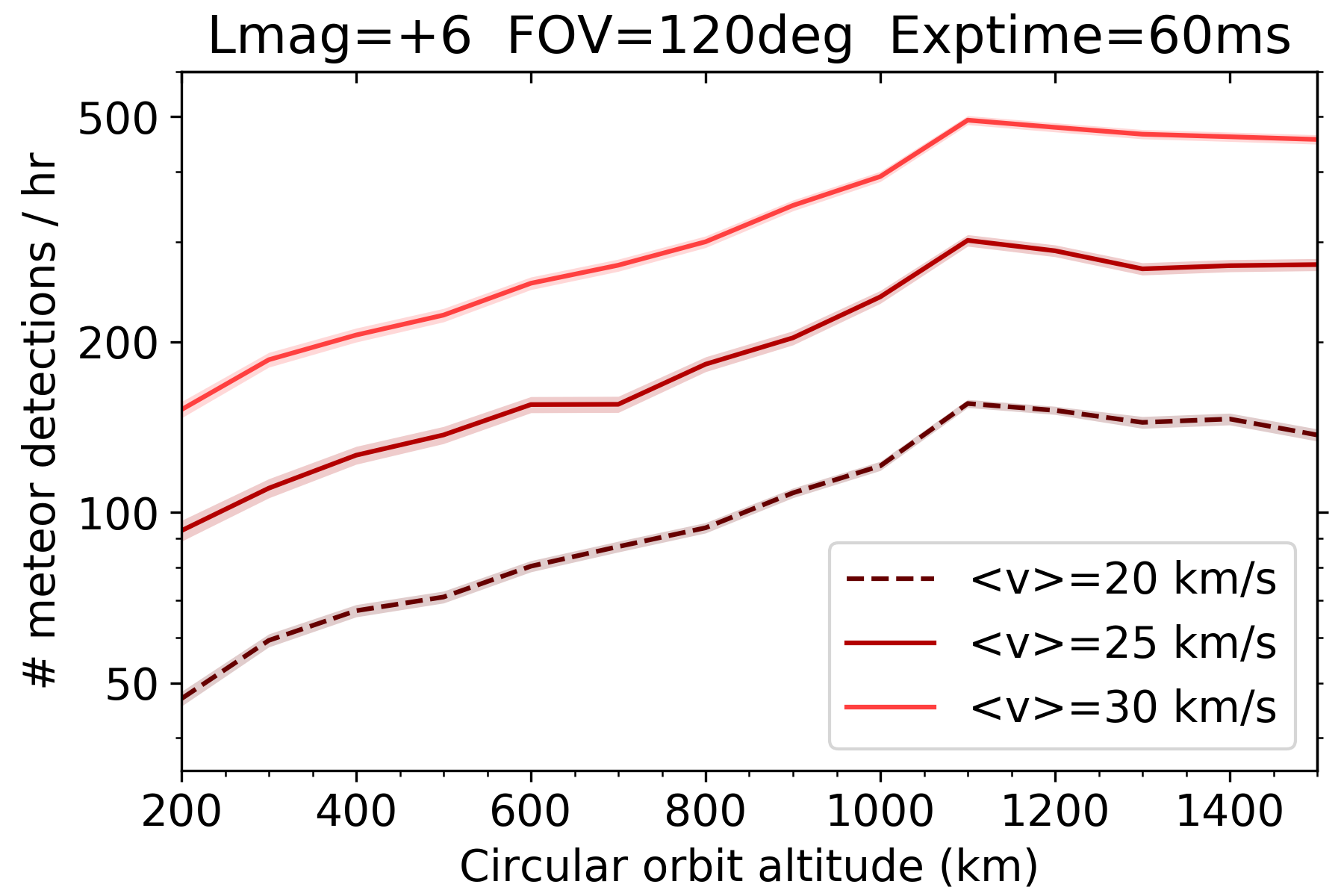}
\includegraphics[width=0.5\columnwidth]{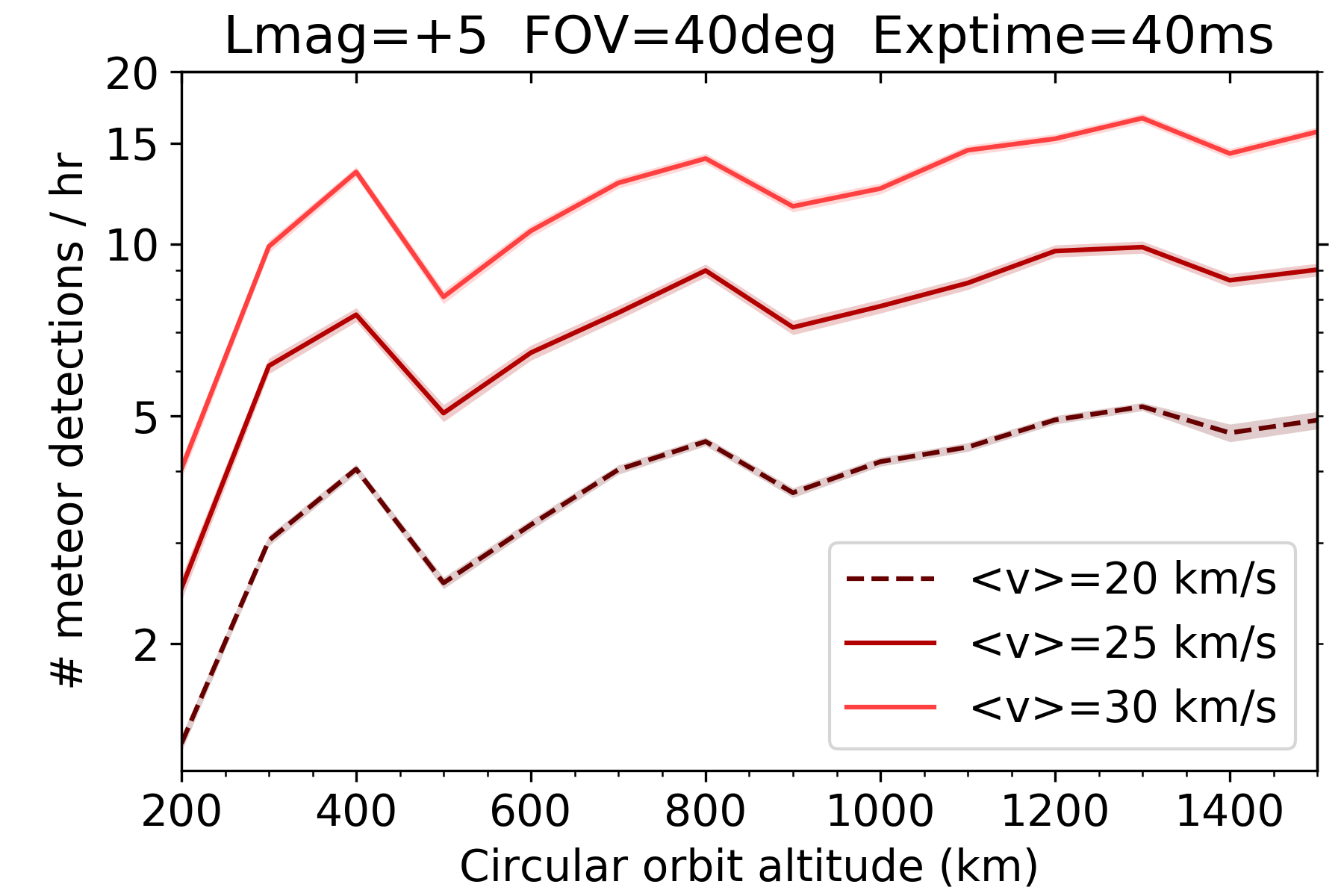}
\caption{Hourly rate of meteor detections at Venus for SPOSH (top) and Mini-EUSO (bottom) for different values of the mode <v> of the speed distribution (Eq.~\ref{eq:speed}) where a brighter colour indicates a faster population of meteoroids.}
\label{FIG:n_vs_altitude_vs_v}
\end{figure}
Increasing the meteoroid speed yields a significant increase in meteor brightness because of the steep dependence of the ablation rate on this parameter (Eq.~\ref{eq:massloss}). We therefore expect the detection rate to also increase. Simulation runs with different values of the mode <v> of the particle speed distribution (Fig.~~\ref{FIG:n_vs_altitude_vs_v}) indeed show an increase by 1.8$\times$ at all altitudes for every incremental change in this parameter by 5 km $\mbox{s}^{-1}$. This finding has implications for the efficient detection of Halley-Type and Long-period Comet meteor showers at Venus, with entry speeds between 50 and 80 km $\mbox{s}^{-1}$ \citep{Beech1998,Christou2010}. Specifically, the corresponding 2- to 3-orders-of-magnitude increase in detection rate for these fast meteors should compensate for their relatively inefficient detection from low altitude orbits due to generally shallow mass distributions.

\section{Conclusions and Discussion}
This work investigated the feasibility of detecting meteors in the atmosphere of Venus from an orbiter. By applying a physical model of the production of meteor light during the atmospheric ablation of a meteoroid we showed that Venus meteors would be moderately brighter, shorter-lived and appear higher in the atmosphere than Earth meteors, the principal cause for these differences being the different density scale heights of the atmospheres. We then carried out orbital survey simulations using the SWARMS s/w, showing that such surveys would yield a higher (by $1.5\times$--$2.5\times$) rate of detections at Venus over Earth. That fact, and the relative accessibility of Venus compared to other solar system planets or satellites with an atmosphere make this planet an ideal venue to conduct the first non-terrestrial meteor survey. Such a survey holds many benefits, including the significant expansion of the observational sample volume in the solar system for potential detection of interstellar meteors, a topic currently under debate in scientific discussions \citep{PenaAsensio.et.al2024}.

The quantitative estimates drawn from our study are, strictly speaking, valid under the specific assumptions defining detector performance, observation orbit, the characteristics of incoming meteoroids and the atmospheric density model. Therefore further simulations were conducted to quantify the parametric dependence of the outcomes. Our findings indicate that the Venus-Earth detection ratio remains consistent and is largely insensitive to variations in the chosen observation orbit and detector characteristics.

{\it EnVision}, an orbital mission to Venus developed by the European Space Agency \citep{Wilson.et.al2022,Widemann.et.al2023}, will perform high-resolution radar mapping and atmospheric observations from a near-circular 220$\times$570 km science orbit\footnote{\tt https://www.esa.int/Science\_Exploration/Space\_Science/EnVision\_factsheet} that closely resembles the baseline 400-km circular orbit in our simulations. Our simulation results can therefore be directly used to quantify the performance of a hypothetical meteor camera onboard that spacecraft.  

For a rough estimate of the detection rate  at Venus, let us assume an instrument with the same performance characteristics and duty cycle as Mini-EUSO and further require that a survey should, at a minimum, aim to yield a sufficient number of detections to measure the population indices - in other words, the relative abundance of faint over bright meteors - of the sporadic background and of the principal showers at Venus. \citet{Marcelli.et.al2023} report a Mini-EUSO detection tally of 24,000 meteors over 90 12-hr sessions on the ISS or $\sim$22 meteors $\mbox{hr}^{-1}$ which we convert to a Venus detection rate of 33--55 meteors $\mbox{hr}^{-1}$ using our findings. If we further assume that (i) shower meteors comprise $\sim$1/3 of the total influx of detectable meteors \citep{Jenniskens.et.al2016}, (ii) a sample of 50 or more meteors per shower - somewhat more than the 29 Leonids detected from space in 1997 \citep{Jenniskens.et.al2000} - is sufficient to estimate the shower population index, and (iii) there exist $\sim$10 strong showers at Venus, we estimate that a survey would require to detect $\sim$10$\times$50+1000 = 1500 meteors over a period of one Venus year ($\sim$225d), an average rate of 0.28 meteors $\mbox{hr}^{-1}$. This exceeds our projected detection rate at Venus by $\gtrsim$2 orders of magnitude and shows that even a less capable version of the instrument - for instance, one featuring a brighter limiting magnitude, a longer single-frame exposure time or a lower duty cycle - will comfortably achieve the survey objectives.

An added complication at Venus is that the data return strategy adopted for Mini-EUSO at the ISS, where the full dataset is stored in memory cards that then are physically returned to Earth for analysis, is not feasible at Venus requiring the availability of appropriate on-board processing to identify and store the relatively small fraction of camera footage containing meteor events for later transmission to Earth. In this context, we note that efficient algorithms for real-time capture and storage of meteor footage have been in use by the meteor community for many years \citep{Yamamoto2005,AtreyaChristou2008,BlaauwCruse2012,Guennoun.et.al2019}. These may serve as suitable starting points to develop the needed capability. 

\section*{Declaration of competing interests}
The authors have no known competing financial interests or personal relationships that could have appeared to influence the work reported in this paper.

\section*{Data availability}
Data will be made available upon reasonable request.

\section*{Acknowledgements}
We thank the two anonymous reviewers whose critical reading of the manuscript improved the presentation of this work. Astronomical research at the Armagh Observatory \& Planetarium is grant-aided by the Northern Ireland Department for Communities (DfC). We are grateful to the Academy of Finland and the Finnish Geospatial Research Institute for supporting the project no.~325806 (PlanetS), which facilitated the development of the modelling approaches utilized in this paper. The program of development within Priority-2030 is acknowledged for supporting the research at UrFU. We thank Dr.~Alexis Bouquet for valuable discussions during the development of his Master Thesis and for sharing the earlier version of the SWARMS code, which proved instrumental in implementing our programme of simulations.

\bibliographystyle{cas-model2-names}
\bibliography{Venus-Meteors-Icarus}
\end{document}